\documentclass[11pt,a4paper]{article}

\usepackage[T1]{fontenc}
\usepackage[utf8]{inputenc}
\usepackage[a4paper,top=3.2cm,bottom=3.6cm,left=3.05cm,right=3.05cm]{geometry}

\usepackage{newpxtext}
\usepackage{newpxmath}
\usepackage[varqu]{inconsolata}
\usepackage{microtype}

\usepackage[authoryear,round]{natbib}
\usepackage{graphicx}
\usepackage{booktabs}
\usepackage{longtable}
\usepackage{xurl}
\usepackage{xcolor}
\definecolor{accent}{HTML}{10099F}% HI blue, links only

\usepackage{titlesec}
\titleformat{\section}{\Large\bfseries}{\thesection}{0.7em}{}
\titleformat{\subsection}{\large\bfseries}{\thesubsection}{0.7em}{}
\titleformat{\subsubsection}{\normalsize\bfseries}{\thesubsubsection}{0.7em}{}
\titlespacing*{\section}{0pt}{1.6\baselineskip}{0.7\baselineskip}
\titlespacing*{\subsection}{0pt}{1.2\baselineskip}{0.5\baselineskip}

\usepackage{fancyhdr}
\fancypagestyle{plain}{\fancyhf{}\fancyfoot[C]{\small\thepage}}

\usepackage{authblk}

\usepackage{hyperref}% orcidlink loads hyperref too, so pass options via hypersetup
\usepackage{orcidlink}
\hypersetup{colorlinks=true,linkcolor=accent,citecolor=accent,urlcolor=accent,
  pdftitle={Curated retrieval versus open web search in public AI
    information services: a coverage-trust trade-off},
  pdfauthor={Hafsteinn Einarsson, Hafsteinn Birgir Einarsson,
    Jon Gunnar Olafsson, Jon Gunnar Thorsteinsson}}

\newcommand{\sep}{\ \textperiodcentered\ }
\graphicspath{{figs/}}

\defcitealias{zhou2024reliable}{L.~Zhou et~al., 2024}
\defcitealias{zhou2024ragtrust}{Y.~Zhou et~al., 2024}

\title{\LARGE\bfseries Curated retrieval versus open web search in public AI
  information services: a coverage–trust trade-off}

\author[1]{Hafsteinn Einarsson\,\orcidlink{0000-0001-5072-3678}\thanks{Corresponding author: \href{mailto:hafsteinne@hi.is}{\texttt{hafsteinne@hi.is}}}}
\author[2]{Hafsteinn Birgir Einarsson\,\orcidlink{0000-0001-9623-487X}}
\author[2]{Jón Gunnar Ólafsson\,\orcidlink{0000-0002-3018-1012}}
\author[3]{Jón Gunnar Þorsteinsson}
\affil[1]{Faculty of Industrial Engineering, Mechanical Engineering and Computer Science, University of Iceland, Reykjavík, Iceland}
\affil[2]{Faculty of Political Science, University of Iceland, Reykjavík, Iceland}
\affil[3]{The Icelandic Web of Science, University of Iceland, Reykjavík, Iceland}
\date{\vspace{0.4em}\small Preprint, July 2026}

\begin{document}
\maketitle
\thispagestyle{plain}

\begin{center}
\begin{minipage}{0.92\textwidth}
\small
\noindent\textbf{Abstract}\par\smallskip
\noindent
Public institutions increasingly use large language models (LLMs) to answer citizens' questions, often pairing a curated knowledge base with live web search, yet whether the sources behind these answers can be trusted has received little empirical scrutiny. We report a pre-launch expert evaluation of Evrópuvefur, an independent, government-funded service run by the University of Iceland that answers questions about the European Union, conducted as Iceland prepared for its referendum of 29 August 2026 on whether to resume EU accession talks. Five domain experts produced 551 evaluations of 449 AI-generated answers, scoring each against a seven-criterion quality rubric and, separately, flagging individual cited sources. We compared two retrieval paths: a curated local corpus (RAG) and open web search. In more than a third of the reviewed web-search answers (35\%, 65 of 187), at least one cited source was flagged, almost always as untrustworthy or irrelevant; curated sources were flagged far less often and only for being out of date. Web search answered more questions, but at the cost of source quality; the curated corpus was trustworthy yet limited in coverage, and the model declined to respond when it fell short. The citation mix also passed over strong sources: across all 287 web-search answers, the system never cited RÚV, the public broadcaster and the country's most widely used news source. A companion prompt ablation shows how weak prompt-level steering is: a trusted-domain list in the system prompt raised the share of citations to listed domains only from 12\% to 21\%. Fluency and topical fit did not predict source trustworthiness. We argue that source trustworthiness is a measurable yet largely invisible dimension of information quality in public AI services, and we discuss transparency-oriented responses and their trade-offs.
\par
\end{minipage}
\end{center}
\vspace{1.5em}

\section{Introduction}
%======================================================================
Public bodies have begun using large language models (LLMs) to answer citizens' questions directly, and generative AI is already in widespread, if uneven, use across government \citep{bright2024,oecd2024}. If the aim is to ground answers in real material rather than the model's own parametric memory, at least two approaches are available: retrieval-augmented generation (RAG) over a curated knowledge base the institution controls \citep{lewis2020}, and live web search over the open internet. We study these two paths separately. The appeal of either is plain. One system can answer far more questions than a hand-written FAQ, in the citizen's own language, at any hour.

Some risks of putting LLMs in front of citizens are by now familiar. A public-sector chatbot can give confidently wrong advice; New York City's business-help bot was caught telling firms to break the law \citep{nycchatbot2024}, and early systems hallucinated and mishandled citations. Those failures have eased as models matured, and they are not our subject. A subtler risk persists even when the prose is fluent and the citations resolve: the sources the system leans on may not deserve the trust readers place in them. Readers were never reliable judges of online source credibility, long before LLMs entered the picture \citep{metzger2007}, and an LLM removes even the ranked list of links that a search engine once left them to inspect. Recent audits make the concern concrete: across eight AI answer engines, citations to news sources were wrong in most tests \citep{towcenter2025}. Citation behaviour has improved as models have matured, and it remains imperfect even in frontier systems built for the task \citep{onweller2026}; through 2024, reliability did not climb steadily as models scaled \citepalias{zhou2024reliable}, even if the newest systems appear to have narrowed some of these gaps. For public information this matters more than for casual search. Citizens often have little choice but to turn to a public service, and whether they trust the information it gives rests on their trust in the institution behind it \citep{belanger2008}; once questioned, that trust can erode into a self-reinforcing spiral of distrust \citep{olafsson2024polarisation}. Someone consulting a public service about a contested political question is not after a plausible paragraph; they are relying on the institution to stand behind the facts and the sources.

Source trustworthiness is, in principle, a dimension of information quality, alongside accuracy, currency (how up to date a source is), and relevance (how well it fits the question asked) \citep{wang1996}. It is also among the easiest to manipulate: a journalist recently seeded a fabricated claim on a personal page and watched mainstream AI assistants repeat it as fact within a day \citep{germain2026}, a reminder that a system drawing on the open web can absorb and amplify misinformation.\footnote{We use \emph{misinformation} for the broader phenomenon of false or misleading information, irrespective of intent, rather than the narrower \emph{disinformation}, which implies deliberate deception \citep{wardle2017information}.} Yet public AI services rarely measure trustworthiness. They log queries and answers; they seldom record whether the cited sources would survive expert scrutiny. These gaps tend to be larger where the stakes are higher: in languages with few high-quality resources, and on politically charged topics where unreliable material is more abundant. Whether it is even an AI provider's place to decide which sources count as trustworthy is itself contested, since source scrutiny shades into editorial control and what makes a source trustworthy is unsettled; a lighter alternative would let users constrain which domains a query may draw on. We return to these questions in the discussion.

We study this gap, between benchmark evidence and the sources a public service would actually cite, in a concrete setting. Evrópuvefurinn (``The Icelandic Web on European Affairs'') is an independent, government-funded service, run by the Icelandic Web of Science at the University of Iceland, that answers questions, in Icelandic, about the European Union (EU) and Iceland's relationship with it. It has received funding from the state but operates independently, an arrangement that makes its credibility central to its mandate. The service had lain largely dormant since Iceland's last accession bid stalled in 2013; as the country prepared for a referendum scheduled for 29 August 2026 on whether to resume EU accession talks, it was being reactivated for a renewed wave of public questions. Ahead of that relaunch we ran a pre-deployment comparison of two ways the system could answer: grounded in a curated corpus of vetted Evrópuvefur articles (RAG), or on open web search. The system generated answers under controlled conditions, which experts then reviewed, weighing a controlled source of evidence against an uncontrolled one before either reached the public.

Over roughly eight weeks, from early May to the start of July 2026, five domain experts reviewed the answers. The questions were not real user queries but a fixed set of 287 questions, generated to span the live debate (Section~\ref{sec:methods}); each was answered twice, once in each mode, so the two paths could be compared on identical inputs. Reviewers scored each answer on a seven-criterion quality rubric and, separately, flagged individual cited sources they judged unfit. This produced 551 scored evaluations and 128 source flags.

Three findings motivate the paper. First, source problems were common, and where the system reached the open web the dominant complaint was trust: experts judged most flagged web sources untrustworthy rather than merely off-topic. Second, the two retrieval paths embodied a coverage--trust trade-off. Web search answered the question far more often, because the open web nearly always has something on point, but it frequently rested on sources experts would not endorse. The curated corpus was trusted by construction yet often lacked coverage, and when it did the model said so rather than inventing an answer. Third, source trustworthiness was largely decoupled from answer quality: once we set aside whether an answer addressed the question at all, most of the gap between the modes closed, and within the web path an answer's fluency and topical fit carried no signal about whether its sources were sound.

Our aim is to make this risk visible and measurable, not to fix it, on the premise that an institution cannot govern or disclose what it does not first measure. We treat source trustworthiness as an information-quality dimension that public AI services can and should assess, we show one practical way to assess it, and we weigh responses ranging from provenance disclosure to expert source labelling against their costs. We are deliberately cautious about remedies that restrict which sources a system may use, because curation by a public body carries its own risks to open access to information.

Concretely, we ask how trustworthy the sources cited by the web-search path are, as judged by domain experts; how the two retrieval paths compare on answer quality and on the reasons reviewers flagged their sources, including what the reviewers' written comments reveal; whether steering the model toward trusted domains in its prompt changes what it cites, which we test with a controlled prompt ablation; and we take up, in the discussion rather than the results, what these patterns imply for the governance and transparency of public AI information services.

The paper makes three contributions. It provides expert-evaluated evidence on the trustworthiness of the sources a public AI information service would cite, in a low-resource language and a high-stakes civic setting. It introduces a simple, reusable review instrument that separates answer quality from per-source trustworthiness. And it frames source trustworthiness as a measurable information-quality dimension for public AI, with a transparency-oriented discussion of what institutions might do about it.

%======================================================================
\section{Background}
%======================================================================
\subsection{AI and large language models in the public sector}
Public administrations have moved quickly to put LLMs in front of citizens. A 2024 survey of UK public-sector professionals found generative AI already in widespread, if disorganised, use, with little governing guidance \citep{bright2024}, and cross-government reviews report rapid, uneven adoption alongside shared concerns about accuracy and oversight \citep{oecd2024}. Independent audits bear the concern out: testing leading LLMs on tens of thousands of public-service questions, the Open Data Institute found inconsistent and sometimes inaccurate answers, and a recurring failure to signal what the model did not know \citep{odi2026}. Conversational interfaces are a common entry point, and a growing body of e-government research examines government chatbots: which social characteristics citizens prefer \citep{ju2023}, whether a chatbot should present a civil-servant identity \citep{liwang2024}, how chatbots reshape public-service provision and public value \citep{larsen2024}, and how citizens and front-line officers weigh their design \citep{hemesath2024}. Because capabilities shift quickly, evidence on any specific model dates fast, so we weight recent audits and tie our results to the systems and period we studied. The move from retrieval to generation also changes the risk profile: a search engine returns ranked links a user can inspect; an LLM returns a single composed answer whose provenance can be harder to see.

\subsection{Information quality and trust in public information}
Whether public information can be trusted is, at root, a question of information quality. Established frameworks treat quality as multi-dimensional, with accuracy, completeness, currency, relevance, and believability among the dimensions information consumers actually care about \citep{wang1996}. Source trustworthiness is one such dimension: an answer can be accurate in its wording yet rest on a source a reader would not accept. In e-government, a citizen's use of a service depends on trust in the institution behind it \citep{belanger2008}, so a service that cites questionable material risks more than a single bad answer: it risks the institutional standing on which its usefulness depends.

What counts as a trustworthy source is not merely a matter of accuracy. It is useful to distinguish edited, mainstream (or legacy) journalism from the alternative media and online platforms (blogs, opinion columns, partisan or advocacy sites). Mainstream outlets are bound by editorial standards, and their role is to cover the different sides of an issue and to act as a watchdog or ``fourth estate'': to hold those in power to account and to inform the public. The function of alternative outlets is more commonly to express positions and specific viewpoints rather than to disseminate traditional news reporting \citep{olafsson2024polarisation}. That content can be a legitimate and valuable part of debate in an open democratic society, but it is not the same as an editorially vetted source for factual claims. This distinction is sharper in a small media market, where a thin and fragile press is more easily dominated and where outside actors can more readily shape the agenda \citep{olafsson2021superficial,omarsdottir2024iceland}; these conditions raise the stakes of which sources a public AI service amplifies.

These risks are not uniform across languages: audits of chatbots verifying political claims find that accuracy varies by language and is weaker outside high-resource ones \citep{kuznetsova2025}, which matters for a service operating in Icelandic, a language spoken by only about 370{,}000 people and correspondingly thin in high-quality machine-readable text. Models are improving in Icelandic, but they still do better when the same task is posed in English than in Icelandic \citep{einarsson2026mazeeval,einarsson2026crosslingual}, so a public service tool providing answers in a smaller language starts from a comparative disadvantage, one that plausibly extends to other low-resource languages in a similar position.

\subsection{Retrieval-augmented generation, provenance, and source credibility}
RAG was introduced to ground model output in retrieved documents and reduce unsupported claims \citep{lewis2020}. Grounding helps, and outright hallucination has grown less common in capable models, though it has not disappeared \citep{ji2023}. Fabricated and miscited references, once rife \citep{walters2023,towcenter2025}, are likewise diminishing as leading models and answer engines improve at attribution, even if they remain imperfect \citep{onweller2026}. We treat these as receding problems and concentrate on one that does not recede with scale: even a system that cites flawlessly still has to draw on sources that are themselves trustworthy. This is where the two retrieval paths diverge. A curated corpus lets a service trust its sources but cannot cover every question; open web search can find material for almost any question but inherits the open web's uneven quality. Grounding, in other words, moves the problem rather than removing it. The question becomes whether the sources a system retrieves and cites would survive scrutiny, and, before that, what we even mean by a trustworthy source, a point we take up in the discussion. Judging the credibility of a web source is, in any case, a skill readers have always applied unevenly \citep{metzger2007}, and surveys of RAG trustworthiness note that the field has emphasised accuracy and efficiency while the trustworthiness of retrieved sources stays underexamined \citepalias{zhou2024ragtrust}.

\subsection{The gap}
Two research gaps follow from this literature. First, most evidence on LLM reliability comes from benchmark prompts rather than the sources a public service would actually cite when answering questions in its domain. Second, few studies cover low-resource languages or high-stakes civic moments such as referenda, where trustworthy material is scarcer and unreliable material more abundant, as the surge of hyperpartisan and misleading content around the 2016 Brexit referendum illustrated \citep{bastos2019brexit,marshall2018posttruth}. We address both gaps by evaluating, before deployment, the quality of the sources that an Icelandic-language public service cites when answering a broad set of questions about a national referendum.

%======================================================================
\section{Case and context}
%======================================================================
\subsection{Evrópuvefur and the 2026 referendum}
Evrópuvefurinn is an Icelandic-language information service about the European Union, run by the Icelandic Web of Science at the University of Iceland. Its remit is to give the public even-handed, sourced answers about the EU and Iceland's ties to it. The service has received state funding (it ran on public funding in 2011--2013 and received a smaller grant again in 2026) but is editorially independent, and that independence is part of what it offers: answers that are not steered by the government of the day. It was opened in 2011, around Iceland's earlier accession bid (Iceland applied to join the EU in 2009 and suspended negotiations in 2013), to explain that process to the public, and it then lay largely dormant for a decade.\footnote{See the service's own account of its origins: \url{https://www.evropuvefur.is/svar.php?id=70881}.}

The EU question returned to the agenda in 2026. The 2024 Alþingi election returned a new centre-left government, the first administration not opposed to EU accession since 2013 \citep{einarsson2024election}, which put the question back on the agenda. The government proposed the vote in March \citep{govis2026referendum}, and on 28 May 2026 the Alþingi (Iceland's national parliament) approved a national referendum, scheduled for 29 August 2026, asking whether Iceland should resume accession negotiations with the EU \citep{althingi2026resolution}. The vote would not decide membership; it would decide only whether talks, dormant since 2013, should restart. Generative AI sharpens the information environment around such a vote. Evidence from elections elsewhere shows the mechanism, though none of it concerns Iceland directly: by the 2024 cycle, language models could already produce election misinformation that readers could not distinguish from genuine material \citep{williams2025}, even if models varied in how readily they complied with such prompts \citep{schlicht2024}. The technology has matured since: today's models write more fluently and refuse such prompts more consistently. What has not been settled is whether the material these systems retrieve and cite can be trusted, the risk a public service must manage when it answers EU-related questions in a charged referendum period.

\subsection{The system}
The service answers a question with an LLM grounded in retrieved evidence. In RAG mode it draws on the Evrópuvefur archive: 742 expert-written, editorially approved answers by 87 contributors across 28 topic areas, the largest being EU affairs, produced between 2011 and 2013 during Iceland's earlier accession process and not updated since \citep{evropuvefurstats}. About 670 EU-relevant answers form the curated corpus. Its age matters for what follows: it is why curated sources were flagged for staleness, and it is part of the coverage gap, since questions about 2026-specific developments often fall outside a corpus frozen in 2013. Questions and articles are embedded with \texttt{multilingual-e5-large}, a multilingual encoder that performs strongly on the Massive Text Embedding Benchmark \citep{muennighoff2023mteb}, and the closest articles are retrieved by approximate nearest-neighbour search. Answers are generated with Google's Gemini models (a Pro model for most queries, a Flash model as fallback), which were among the best-performing models in Icelandic at the time of the study,\footnote{Judged against the public Icelandic LLM leaderboard maintained by Miðeind \citep{mideind_leaderboard}.} with the retrieved articles supplied as context and cited in the answer.

For this evaluation the system answered each question in one of two modes, recorded for every query. In \emph{RAG} mode the model is grounded in the curated local corpus described above. In \emph{web-search} mode the model runs its own web searches, with full control over the queries it issues and the pages it draws on, and cites external web pages. Its instructions explicitly steered it toward reputable, primary, and authoritative sources (a guiding list of domains that \texttt{esbvaktin.is} classifies as ``high confidence'', which the prompt asks the model to prefer but which is not enforced as a hard constraint; Section~\ref{sec:methods}) and away from blogs, opinion columns, and partisan sites; we give both modes' prompts verbatim in Appendix~\ref{app:prompts} and return to the effect of that instruction in the discussion. The two modes give a direct contrast between a controlled source of evidence, which the institution has vetted, and an uncontrolled one drawn from the open web. This contrast is the backbone of the analysis.

%======================================================================
\section{Methods}
\label{sec:methods}
%======================================================================
\subsection{Question generation}
Because the service was not yet public, we could not draw on real user queries, so we generated a broad, self-contained question set designed to span the public debate. The source material came from \texttt{esbvaktin.is}, chosen because it is an open, nonprofit, and transparent EU-referendum fact-checking project built independently by a doctoral student at the University of Iceland (see Acknowledgements), not an official University project: it tracks the public discussion around the referendum, collects parliamentary speeches, fact-checked claims, evidence records, and media reports, and classifies the domains it draws on by confidence level, a ready-made and documented classification we could reuse to guide source selection in web-search mode. We weigh the risk that this introduces (an LLM-assisted aggregator shaping our own pipeline) in Section~\ref{sec:limitations}. We embedded passages from this corpus and clustered them (UMAP for dimensionality reduction, then HDBSCAN) to surface the recurring themes of the debate. Within each cluster we chose seed passages by maximal-marginal-relevance selection: a greedy procedure that builds a set one item at a time, each time adding the passage that scores highest on a weighted combination of relevance to the cluster and dissimilarity to the passages already chosen. Intuitively, it avoids picking several near-identical passages, so the seeds cover a theme broadly rather than restating its single most typical point. A generation model (Google Gemini~3 Pro) wrote candidate questions from each seed; a second model checked that each question stood on its own without referring back to a source text, and near-duplicates were removed by embedding similarity. The result was 287 questions spanning six types: policy reasoning, comparative cases (for example Norway or the European Economic Area, EEA), competing discourse positions, historical context, the referendum itself, and common misconceptions. They were deliberately weighted toward reasoning rather than trivia, to mirror the questions actually circulating in the debate.

\subsection{Data}
The 287 generated questions, answered in both modes, produce 574 eligible answers, evenly split (287 RAG, 287 web search); queries logged during development and testing, and answers generated after the study window for a separate batch of questions as the service moved toward production, are excluded. Five reviewers produced 551 evaluations covering 449 distinct answers (262 RAG, 187 web search) and 128 source flags. They also edited 103 answers and marked them ready for publication; whether any of those is formally published is a later editorial decision, taken outside this study. Reviewers worked through a shared queue rather than a balanced assignment, so the two modes were reviewed in unequal numbers, a limitation we return to in Section~\ref{sec:limitations}. For the 179 questions reviewed in both modes the two answers were produced from the same underlying question, each mode wrapping it in its own prompt. This matched subset lets us compare the modes while holding the question fixed, which controls for question difficulty: we use McNemar's test for the binary ``answers the question'' criterion and the Wilcoxon signed-rank test for the composite score. We do not run a paired comparison of source flags, because the available flag reasons differ by mode (Section~\ref{sec:methods}) and so are not comparable across the pair.

\subsection{Expert review instrument}
Five domain experts reviewed AI-generated answers through a dedicated review interface. The instrument has two independent parts.

The first is a \emph{quality rubric} of seven yes/no criteria applied to the answer as a whole: whether it answers the question, is factually accurate, draws on relevant sources, is free of hallucinations, keeps to an appropriate scope, reads well in Icelandic, and is publishable with only minor edits. We report the per-criterion pass rates and summarise the rubric as a composite score, the number of criteria an answer passes, from 0~to~7.

The second is a \emph{source-flagging} mechanism that operates on individual cited sources rather than the answer as a whole. A reviewer who judged a cited source unfit flagged it and recorded a reason, and the available reasons differed by mode, by design. Curated local articles had already passed editorial vetting, so reviewers assessed them only for currency: the one applicable reason was \emph{outdated}. Web pages, which entered an answer with no prior vetting, were assessed instead for whether they were \emph{untrustworthy} or \emph{irrelevant}. This asymmetry reflects the question we set each path. For the curated corpus, has the vetted material aged? For the open web, did the system reach material worth citing at all? It also means flag \emph{reasons} are not directly comparable across modes, which we keep in view when interpreting them. Reviewers could add a free-text comment, and most did (90 of 128 flags). We analyse those comments in Section~\ref{sec:taxonomy} with the help of an LLM, to see what reviewers actually objected to. Separating per-source flags from the whole-answer rubric is what lets us ask whether a good answer can still rest on a bad source.

\subsection{Analysis}
We report proportions with Wilson 95\% confidence intervals \citep{wilson1927}. We compare modes with the chi-squared test, or Fisher's exact test where expected counts are small, and summarise effect sizes with Cram\'er's $V$ and odds ratios; for the paired subset we use the Wilcoxon signed-rank test. For inter-rater reliability we use Gwet's AC1 as the headline measure \citep{gwet2008}, because it is robust to the high-prevalence problem that distorts Cohen's and Fleiss' kappa when most judgments fall in one category; we report kappa and Krippendorff's alpha \citep{krippendorff2018} alongside it, and read coefficients against conventional thresholds \citep{landis1977}. To categorise the free-text flag comments, we built a fixed set of reason codes by reading all of them, then had an LLM (Gemini~3.5 Flash, with a constrained structured-output schema) assign every applicable code to each comment; a comment can carry more than one. The figures and statistics are reproducible from the evaluation export and analysis code described in the Data availability statement.

\subsection{Review procedure}
Reviewers signed in to a dedicated web interface and were served one query at a time in randomised order, with queries that no one had yet evaluated prioritised so that corpus coverage built up before annotators doubled up on the same items (which is why inter-rater overlap is sparse; Section~\ref{sec:rq-reliability}). Each reviewer saw the question and the full generated answer with its inline citations, scored the seven rubric criteria, and flagged any cited source in the same pass. The interface presented the rubric items and flag reasons in English (the flag labels were ``Outdated'', ``Irrelevant'', and ``Not trustworthy''); the answers under review, and the reviewers' free-text comments, were in Icelandic. Because the cited sources were visible (local article links in RAG mode, external URLs in web-search mode), the retrieval mode was not blinded; this is an unavoidable feature of source review, since judging a source means seeing it, and we note it as a possible influence on judgments. Reviewers could not see one another's evaluations.

\subsection{Prompt-ablation experiment}
\label{sec:methods-ablation}
To isolate what the trusted-domain list in the web-search prompt actually does, we ran a controlled prompt ablation, an experiment that removes one component of a system, here the trusted-domain list, to measure what that component contributes. It ran alongside the expert review, in April 2026. Each of the 287 study questions was answered twice more by the production setup (Gemini~3 Pro through the same API route and web-search plugin the deployed system uses), with one difference in how citations are captured: instead of a free-text answer whose reference list we parse, the ablation constrains the model to a structured output with an explicit citation list. The two arms are: once with the full production-style prompt including its trusted-domain list, and once with the ``Source selection'' section stripped, everything else identical. Of the 574 requests, 549 completed (275 with the list, 274 without); failed requests returned no answer and drop out of both arms. Redirect URLs returned by the search tool were resolved to their target pages before analysis, and every citation was classified as \emph{on-list} (its host matches a listed domain exactly or is a subdomain of one) or \emph{off-list}. This classification measures compliance with the list, not trustworthiness: an off-list citation is not necessarily untrustworthy. We also replicate the audience-profile analysis of Section~\ref{sec:rq-profile} on each arm's citations.

\subsection{Ethics and transparency}
The reviewers were five students in the social sciences (three master's students, one PhD student, and one bachelor student, all at the School of Social Sciences, University of Iceland), selected for their subject-matter background in European and EU affairs and recruited with the assistance of a faculty member in that field (see Acknowledgements). They were paid contributors, not anonymous survey participants, and were compensated per item (ISK~1500, about EUR~10.45, per reviewed answer). Under the relevant institutional rules the work did not require formal ethics review. All analysed data were anonymised, reviewers worked independently and could not see one another's evaluations, and they understood that their evaluations would be used to assess and improve the service. The service is funded by Iceland's Ministry for Foreign Affairs but is editorially independent; the funding was administered by one of the authors and the ministry had no role in the evaluation or its reporting.

%======================================================================
\section{Findings}
%======================================================================
Across the 551 evaluations, answers passed a mean of 5.1 of the 7 quality criteria. Our focus is the 128 source flags and what they reveal about where citizens' answers would come from.

\subsection{Source problems are common, and on the open web the complaint is trust}
\label{sec:rq1}
\begin{figure*}[t]
  \centering
  \includegraphics[width=\linewidth]{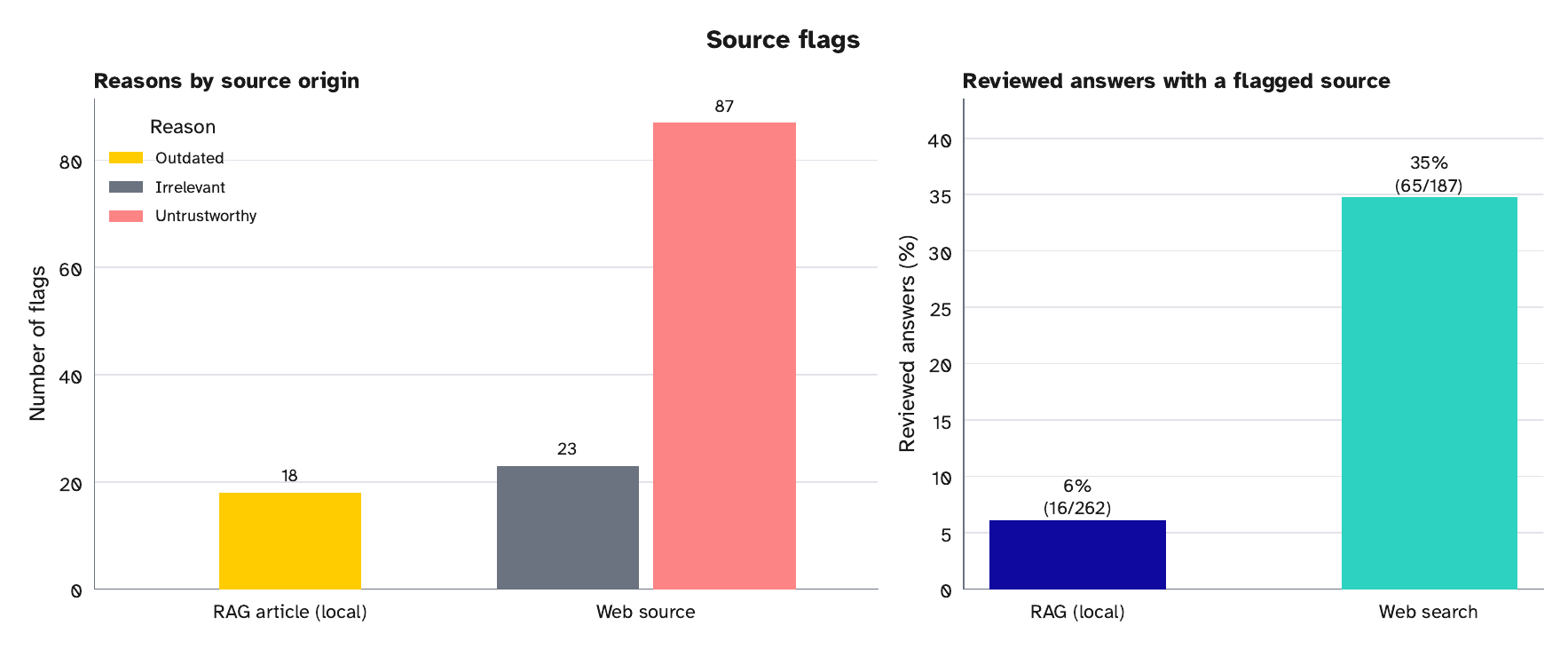}
  \caption{Source flags. (A) Flag reasons by source origin; the reasons
  available to each path differ by design, so this shows the reason distribution
  within each path, not a discovered contrast. (B) Share of reviewed answers with
  at least one flagged source, by mode (web 65/187, RAG 16/262); because the
  available reasons differ by mode, this cross-mode comparison is descriptive.}
  \label{fig:flags}
\end{figure*}

Reviewers flagged 128 cited sources. On the open web, where a source could be flagged as untrustworthy or irrelevant, trust was the dominant complaint: of 110 web flags, 87 were for untrustworthiness and 23 for irrelevance (Figure~\ref{fig:flags}A). The 18 curated-corpus flags were, by the design of the instrument, all for being outdated. Because the two paths were held to different criteria (Section~\ref{sec:methods}), the left panel should be read as the reasons available to each path, not as evidence that the paths fail in categorically different ways. The substantive result is within the web path: when experts could question a web source's reliability, they frequently did, and far more often than they found it merely off-topic.

\subsection{Web search is where the trust risk concentrates}
\label{sec:rq2}
When reviewers examined a web-search answer, 35\% of the time they flagged at least one of its sources (65 of 187 reviewed web answers); for RAG the figure was 6\% (16 of 262) (Figure~\ref{fig:flags}B). Per source the rate is lower but still substantial: across the 187 reviewed web answers the model cited 1{,}088 sources (442 distinct URLs), a mean of 5.8 per answer, so the 110 flags fall on roughly 10\% of cited sources and 25\% of distinct URLs. These rates are not a like-for-like comparison with RAG, since web and curated sources were held to different standards. Because local articles were assessed only for currency, the study can estimate web-source trust problems directly but cannot estimate a like-for-like trust gap between web and curated sources. The web figure nonetheless stands on its own: in more than a third of the web answers an expert examined, a cited source was one they judged untrustworthy or irrelevant.

\begin{figure}[t]
  \centering
  \includegraphics[width=\linewidth]{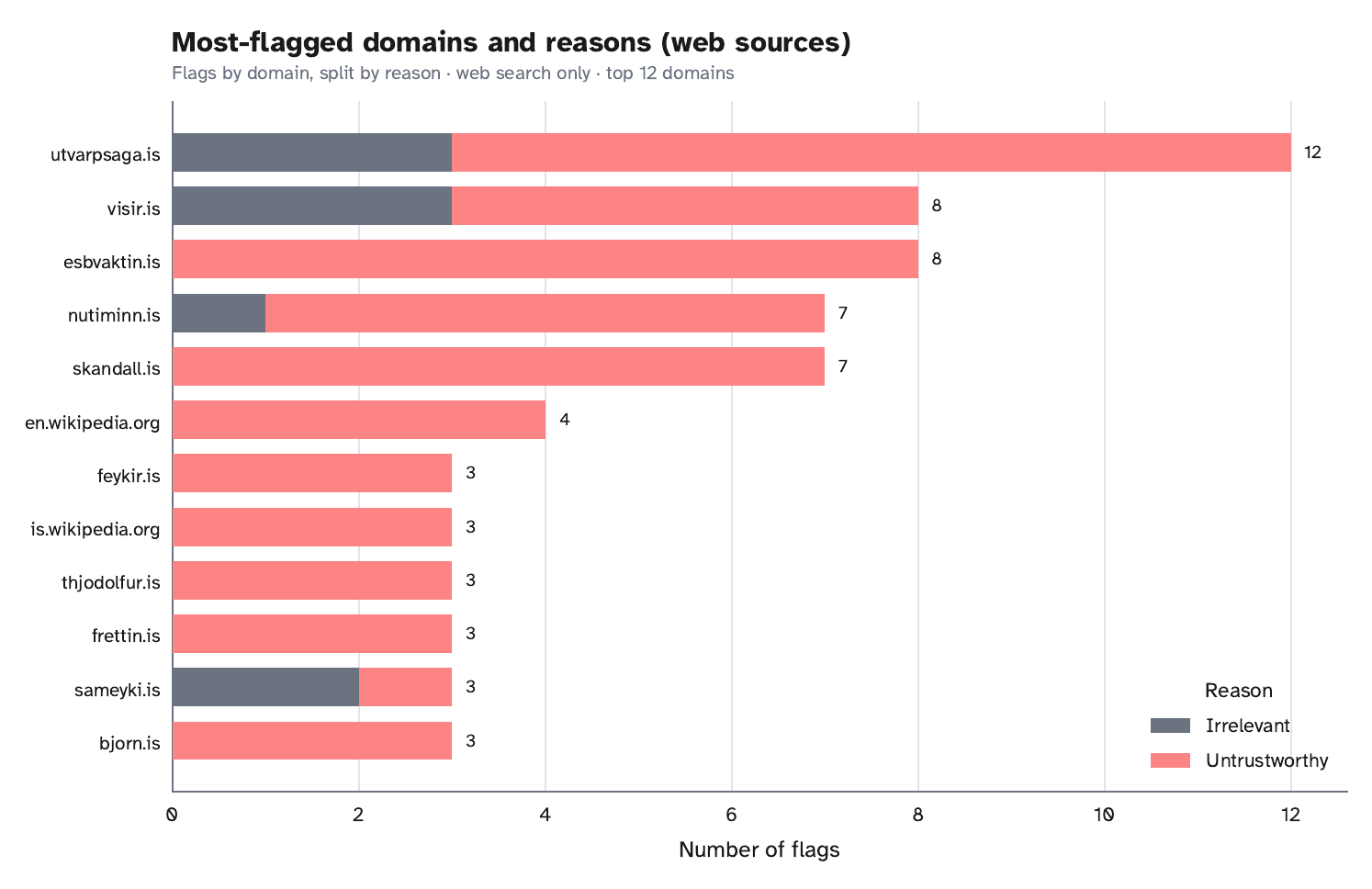}
  \caption{Most-flagged web domains, coloured by reason (web sources only).}
  \label{fig:c1}
\end{figure}

The flagged web sources cluster on a recognisable set of domains (Figure~\ref{fig:c1}), and the flags track source type more than mere frequency. The trusted aggregator \texttt{esbvaktin.is} was the most-cited domain (84 citations) and was flagged on 10\% of them, whereas alternative, opinion-driven and lightly edited outlets were flagged far more often: the pages of one broadcaster with a declared anti-EU stance were flagged on 27\% of their citations, and the most-flagged tabloid-style site on more than half. The contrast tracks the mainstream/alternative-media distinction drawn in the Background rather than mere citation frequency. The problem is less that the web has no relevant pages on these questions and more that the pages the system reached were often ones that experts would not endorse.

\subsection{What reviewers objected to}
\label{sec:taxonomy}
\begin{figure}[t]
  \centering
  \includegraphics[width=\linewidth]{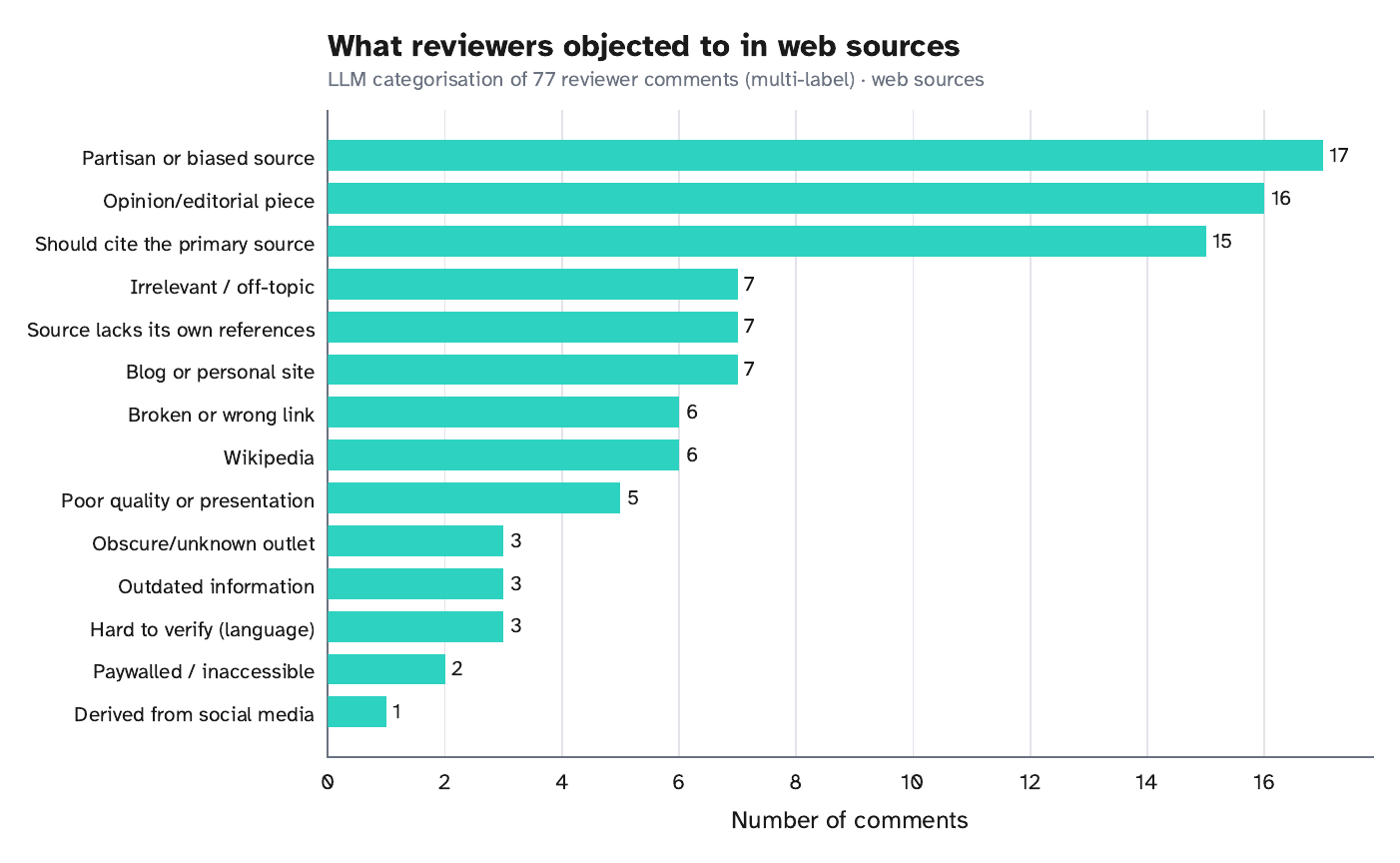}
  \caption{What reviewers objected to in flagged web sources, from an LLM
  categorisation of their written comments (multi-label; 77 web comments).}
  \label{fig:taxonomy}
\end{figure}

The flag reasons (outdated, irrelevant, untrustworthy) are coarse, so we looked at the written comments behind them. Reading the comments (90 of the 128 flags carry one), we built a set of recurring objections and had an LLM tag each comment with every objection that applied (Section~\ref{sec:methods}); the 77 web comments are summarised in Figure~\ref{fig:taxonomy}. Two complaints dominate, and both are about the type of source rather than its topic: the cited page came from a partisan or otherwise non-objective outlet (17 comments), or it was an opinion or editorial piece (16). A second cluster concerns sourcing practice: the answer cited a secondary report when a primary source was available (15), or a page that itself contained no references (7). The rest spread across recognisable web-quality problems: personal blogs (7), Wikipedia (6), broken links (6), visibly poor presentation (5), paywalled or otherwise inaccessible pages (2), and material in a language that made it hard to verify (3). The picture is consistent with the domain analysis: the model reached pages that were on topic but editorially weak. The labelling was LLM-assisted but not left unchecked: the authors read every comment against its assigned codes and corrected the few that were off, and we reproduce all 77 web comments, an English translation, and their final codes in Appendix~\ref{app:flagaudit} so the coding can be audited directly.

\subsection{Trust is decoupled from answer quality}
\label{sec:rq3-context}
\begin{figure*}[t]
  \centering
  \includegraphics[width=\linewidth]{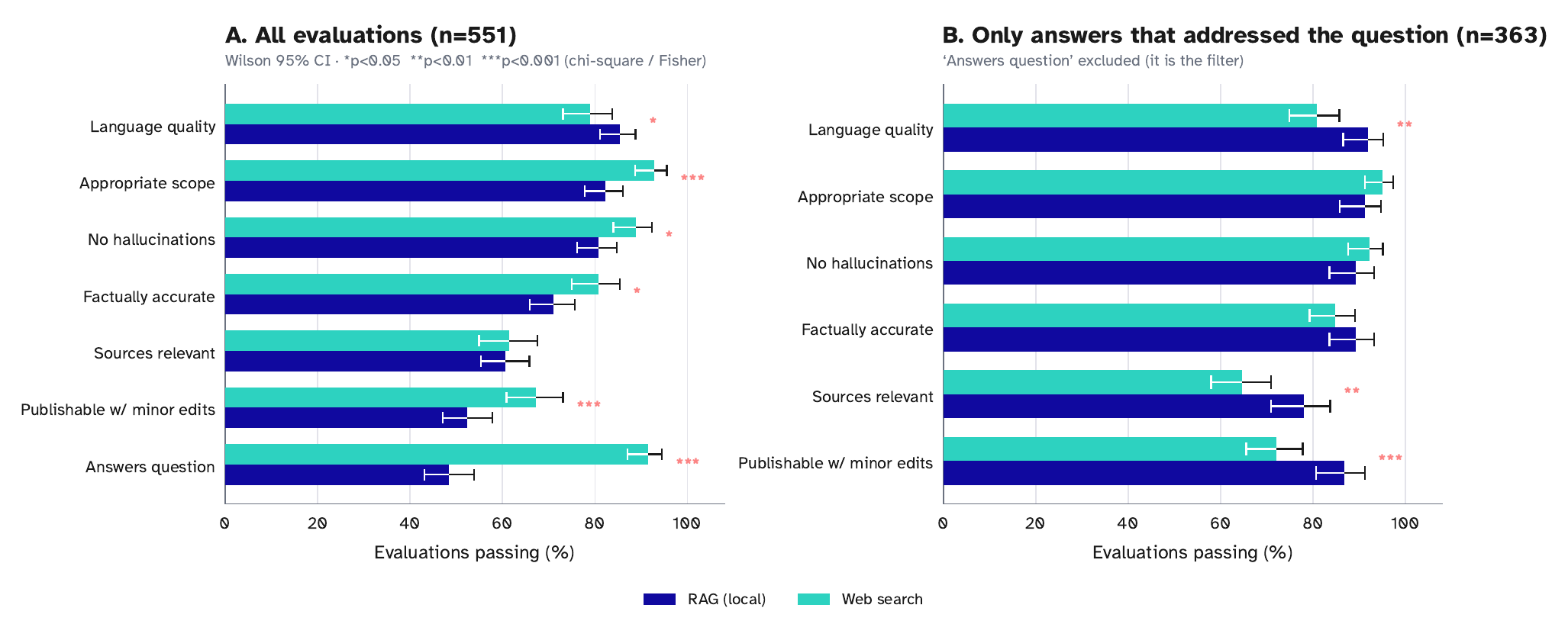}
  \caption{Quality-criterion pass rates by mode, with Wilson 95\% intervals.
  (A) All evaluations. (B) Only evaluations whose answer passed ``answers the
  question'', with that criterion dropped; this panel shows what excluding the
  unanswered questions, RAG's coverage failure mode, does to each remaining
  criterion. Stars mark significant differences within each panel.}
  \label{fig:rubric}
\end{figure*}

A natural worry is that flagged answers are simply bad answers, which an ordinary quality review would already catch. The rubric scores say otherwise. Web answers far outperformed RAG on the single criterion of whether they addressed the question (91.5\% versus 48.5\%; $p<0.001$; Figure~\ref{fig:rubric}A). The matched questions tell the same story: among the 179 questions reviewed in both modes, web answered 82 that RAG did not, against only 8 the other way (McNemar's test, $p<0.001$). The reason is instructive: when the curated corpus lacked material for a question, the model usually said so rather than inventing an answer. RAG answers fell back on a stock disclaimer, in essence ``according to Evrópuvefur's material, there is not enough information to answer this'', which we detected with a keyword pattern and confirmed by reading a sample of the matches. Among RAG answers that failed the ``answers the question'' criterion, 87\% carried such a disclaimer; the phrasing appeared in 70\% of all RAG answers but in only about 1\% of web answers, which almost always returned something on point. The low RAG figure thus reflects the curated corpus's coverage, not the model's writing. Web answers also scored significantly higher on appropriate scope (92.8\% versus 82.3\%; $p<0.001$), publishability with minor edits (67.3\% versus 52.4\%; $p=0.001$), factual accuracy (80.7\% versus 71.0\%; $p=0.010$), and freedom from hallucinations (88.8\% versus 80.8\%; $p=0.012$), gaps that, as the next paragraph shows, are largely carried by the coverage disclaimers. RAG answers scored higher on the language-quality criterion, the reviewers' yes/no judgment that an answer reads well in Icelandic (Section~\ref{sec:methods}), at 85.4\% versus 78.9\% ($p=0.049$); the two modes did not differ on the relevance of their sources.

Because a declined RAG answer mechanically fails not only ``answers the question'' but several of the other criteria with it, the raw cross-mode bars understate the quality of the answers RAG actually produces. Panel~B of Figure~\ref{fig:rubric} therefore recomputes every criterion over only the evaluations whose answer passed ``answers the question'' (159 RAG and 204 web evaluations), so the reader can see, criterion by criterion, what dropping the unanswered questions does to the comparison. It reverses much of it. On this answered-only subset the two modes no longer differ on factual accuracy (89\% versus 85\%), appropriate scope, or freedom from hallucinations (all $p>0.05$), and RAG moves ahead on relevant sources (78\% versus 65\%; $p=0.006$), language quality (92\% versus 81\%; $p=0.003$), and publishability with minor edits (87\% versus 72\%; $p<0.001$); the composite over these six criteria favours RAG at 5.3 versus 4.9 of~6 (Mann--Whitney $p<0.001$). In other words, when the curated corpus does cover a question, the answer RAG returns is typically of high quality; the mode's low aggregate scores are a coverage effect, not a generation deficit. We treat this subset analysis as exploratory, since conditioning on ``answers the question'' selects a different set of questions in each mode.

On the ``answers the question'' criterion, web answers rarely failed (9\%), and the few that did were idiosyncratic rather than systematic. To see how, we read all nineteen failing evaluations (covering eighteen answers) together with the reviewers' notes. Two were outright technical breakdowns: in one the model returned only its own search-and-reasoning scratchpad (``investigating EU fisheries negotiations\ldots'') with no finished answer, and in another, on the legal duties of a parliamentary article, it pasted a raw search-tool trace that the reviewer found unreadable. A second group answered a subtly different question than the one asked, usually because the question carried a false premise or an outdated one: one assumed Iceland had adopted the 2011 draft constitution, which it has not; another asked about safeguarding ``all the nation's resources'' while the answer addressed only fisheries; a third presented as prospective a regulation Iceland had already implemented in 2022. A third group leaned on a single weak source, such as an opinion blog opposed to the referendum, that the reviewer judged too thin to count as answering, and the remainder were factual quibbles in otherwise ordinary answers. None of these were the coverage disclaimers that dominated the RAG failures, so the two modes failed this criterion for entirely different reasons.

Because the ``answers the question'' criterion, driven by coverage, dominated the comparison, we also recomputed the composite over the other six criteria across all evaluations, unanswered ones included. Much of the gap closed. Across all evaluations the modes were statistically indistinguishable (4.33 versus 4.70 of 6; Mann--Whitney $p=0.15$), and a paired test on the 179 questions reviewed in both modes left a residual edge for web (4.75 versus 4.19 of 6; Wilcoxon $p=0.003$), far smaller than the coverage-driven gap on ``answers the question'' itself.

A sharper test stays inside the web path: do answers whose sources were flagged score worse than those whose sources were not? On the criteria that capture how an answer reads, they do not. Web answers with at least one flagged source were statistically no different from unflagged ones on whether they answered the question (90\% versus 93\%), read well in Icelandic (73\% versus 83\%), stayed in scope (90\% versus 95\%), or avoided hallucinations (84\% versus 92\%; all $p>0.05$, Fisher's exact). Two criteria did separate them: ``sources relevant'' (33\% versus 81\%) and the overall ``publishable with minor edits'' judgment (47\% versus 81\%; both $p<0.001$), and the six-item composite was lower for flagged answers (4.0 versus 5.1 of 6; $p<0.001$). The first is close to definitional, since a flagged source is often an irrelevant one; the second is more telling, because the reviewer making the overall publishability call sensed something amiss that the surface criteria did not. So fluency and topical fit carry no signal about whether the sources underneath are sound, while a reviewer's overall editorial judgment carries some. Ordinary answer-level review would catch part of the source-trust problem and miss part of it, which is exactly why source trustworthiness has to be assessed on its own.

\subsection{Coverage and source quality by question type}
\label{sec:rq-category}
\begin{figure*}[t]
  \centering
  \includegraphics[width=0.49\linewidth]{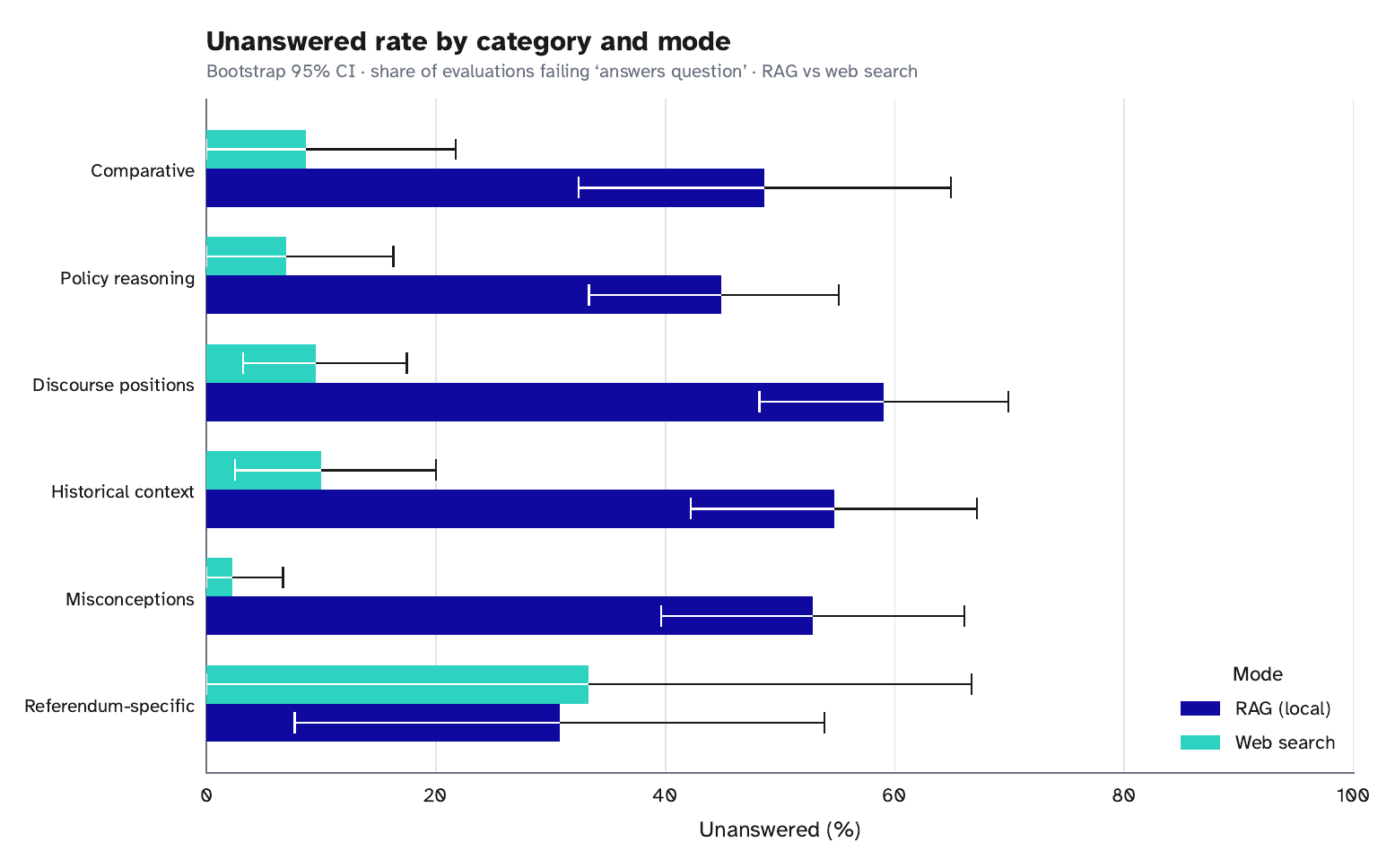}\hfill
  \includegraphics[width=0.49\linewidth]{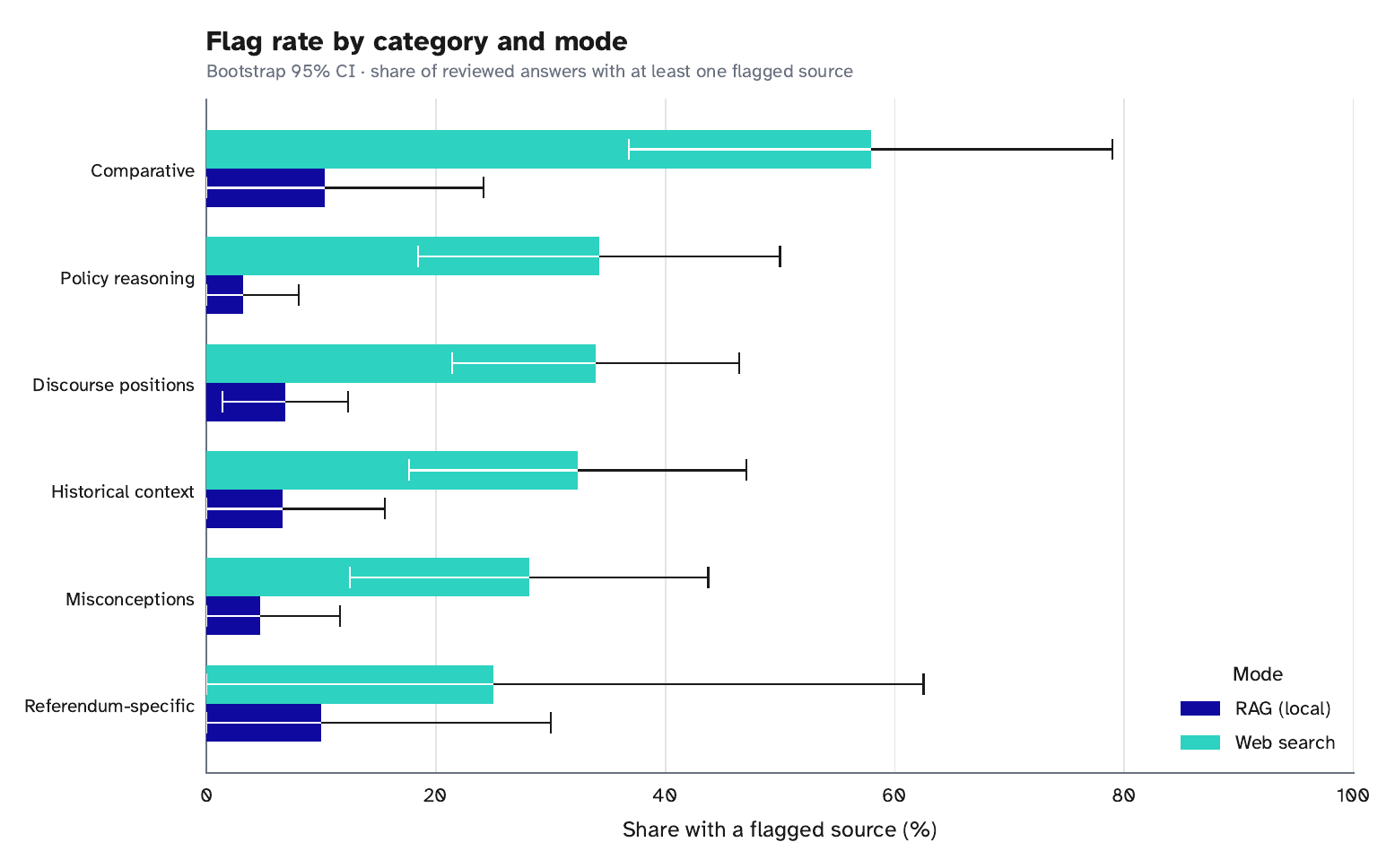}
  \caption{Patterns by question type. (Left) Unanswered rate (share of evaluations
  failing ``answers the question''), by type and mode. (Right) Share of reviewed
  answers with at least one flagged source, by type and mode. Whiskers are
  bootstrap 95\% confidence intervals (10{,}000 resamples); the smaller types
  (comparative, referendum-specific) rest on few reviewed answers, hence the wide
  intervals. RAG = blue, web search = teal.}
  \label{fig:category}
\end{figure*}

The generated question set was not uniform: it spanned six types, weighted toward
the live debate, with discourse positions (83 questions) and policy reasoning (67)
the largest, followed by historical context (49), misconceptions (45),
comparative cases (31), and referendum-specific questions (12). Breaking the two
headline patterns down by type shows that both hold across the board rather than
resting on one kind of question (Figure~\ref{fig:category}).

The coverage gap is visible in every type. RAG left between a third and three-fifths
of questions unanswered depending on type (highest for discourse positions at 59\%
and historical context at 55\%, lowest for referendum-specific questions at 31\%),
whereas web search answered nearly all of them (unanswered rates of 2--10\% across
the five larger types; 33\% for the tiny referendum-specific type, which rests on
nine evaluations). The
pattern matches the coverage story: the curated corpus, frozen in 2013, is thinnest
exactly where the debate has moved on, such as evolving discourse positions. The
source-trust problem is likewise present in every type on the web path, where the
share of reviewed answers carrying a flagged source ranged from 25\% to 58\%,
against at most 10\% for RAG. The web flag rate was highest for comparative cases
(58\%), though the smaller types (comparative,
referendum-specific) rest on few reviewed answers and should be read with that in
mind. No question type was simultaneously well-covered by the curated corpus and
free of web-source flags, which reinforces the coverage--trust trade-off rather
than localising it to one topic.

\subsection{The political profile of the cited sources}
\label{sec:rq-profile}
The expert flags are one lens on source quality; an independent one is who, in
the wider public, actually reads a given outlet. We characterised each Icelandic
news outlet the web-search mode cited using the Fjölmiðlanefnd (Media Commission)
2024 survey, a nationally representative survey of media use and political
orientation \citep{fjolmidlanefnd2024}. Following the audience-polarisation
method of \citet{fletcher2020polarized}, used in the Reuters Institute Digital
News Report \citep{newman2022dnr} and, for Iceland, by
\citet{olafsson2024polarisation}, we placed each outlet on two axes by the
make-up of its audience: a left--right axis and, more salient for an EU vote, a
nationalism--internationalism axis. For each axis an outlet's audience skew is
the share of its readers at the high end minus the share at the low end, read
relative to the national population. We then took every reference the system
cited (parsed from each answer's reference list) across the 287 eligible
web-search answers, kept the 153 whose domain mapped to a surveyed outlet (nine
distinct cited outlets, of eleven surveyed), and computed the citation-weighted
audience profile of those references. Curated (RAG) answers cite only the in-house corpus, so this analysis
necessarily concerns the open-web path.

\begin{figure*}[t]
  \centering
  \includegraphics[width=\linewidth]{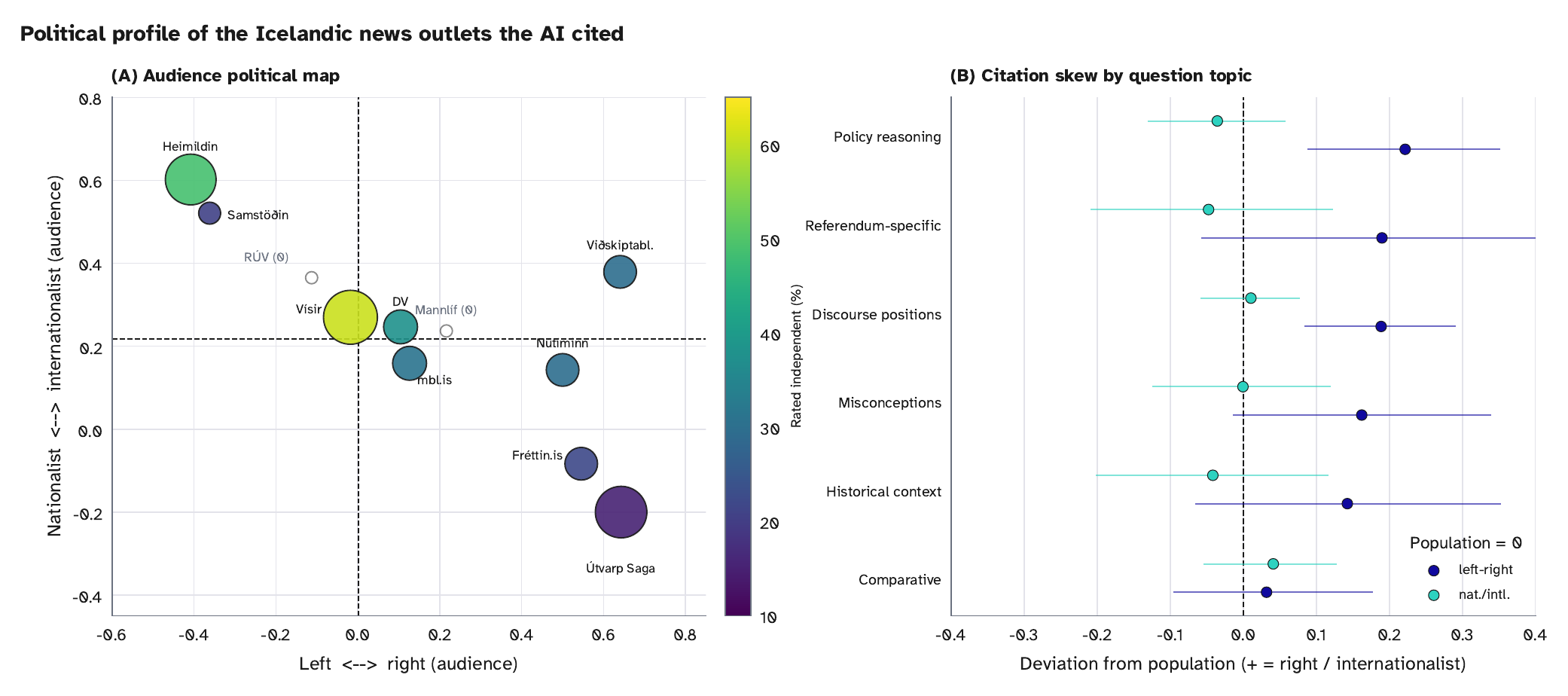}
  \caption{Political profile of the Icelandic news outlets the web-search mode
  cited, from the Fjölmiðlanefnd 2024 survey. (A) Each outlet placed by its
  audience's left--right skew (x) and nationalist--internationalist skew (y);
  bubble size = number of references the system gave it, colour = share of the
  public rating it politically independent, dashed lines = national population.
  Open circles (RÚV, Mannlíf) were not cited. (B) Citation-weighted audience skew
  of the cited references relative to the population, by question topic, on both
  axes (bootstrap 95\% CIs). Positive = right-leaning / internationalist audience.}
  \label{fig:profile}
\end{figure*}

Two patterns emerge (Figure~\ref{fig:profile}). On the left--right axis, the
sources the system cited lean right of the population (citation-weighted skew
$+0.17$ against a population near zero; 57\% of matched references went to outlets
with right-leaning audiences), and this tilt held across every question type,
strongest for policy-reasoning questions. On the nationalism--internationalism
axis the average cited source matched the (internationalist-leaning) population
rather than shifting it, but the only two outlets whose audiences lean
\emph{nationalist}, Útvarp Saga and Fréttin.is, were both among the alternative
outlets experts flagged, and Útvarp Saga drew more flags than any other outlet. The two-dimensional map makes the structure plain:
the heavily flagged outlets sit together in the right-leaning, nationalist,
low-independence corner (independence here is the survey's measure, the share of
respondents who rate an outlet independent of political interests), while the
outlets whose audiences lean internationalist (Heimildin, Samstöðin, RÚV) drew
few or no flags. This is independent corroboration, from public survey data rather than our
own reviewers, of both the expert flags and the mainstream/alternative
distinction of Section~\ref{sec:methods}: the per-outlet audience profile,
perceived independence, citation counts, and flag counts are reported in
Appendix~\ref{app:profile}. The clearest exception is Vísir, a mainstream portal
the public rates as independent (61\%) yet which drew eight flags; reading them
shows a mix of opinion pieces judged untrustworthy and off-topic articles judged
irrelevant, rather than a problem with the outlet as such. This is less
surprising than it first appears: Vísir is the main Icelandic venue for opinion
pieces contributed by the general public, which the site publishes in a section
clearly separated from its news reporting and editorial content, and every
untrustworthiness flag on Vísir fell on such a contributed piece rather than on
the newsroom's output. These survey-based
measures capture an outlet's \emph{audience}, not the slant of any individual
cited article, and cover only the Icelandic-news subset of citations; we treat
them as exploratory and, given small audiences for the niche outlets, report
bootstrap 95\% intervals.
One absence deserves its own sentence: RÚV, the state-funded public
broadcaster and the country's most widely used news source, was never cited
by the deployed system across the 287 web-search answers.

\subsection{Does the source list steer citation behaviour?}
\label{sec:rq-ablation}
\begin{figure*}[t]
  \centering
  \includegraphics[width=\linewidth]{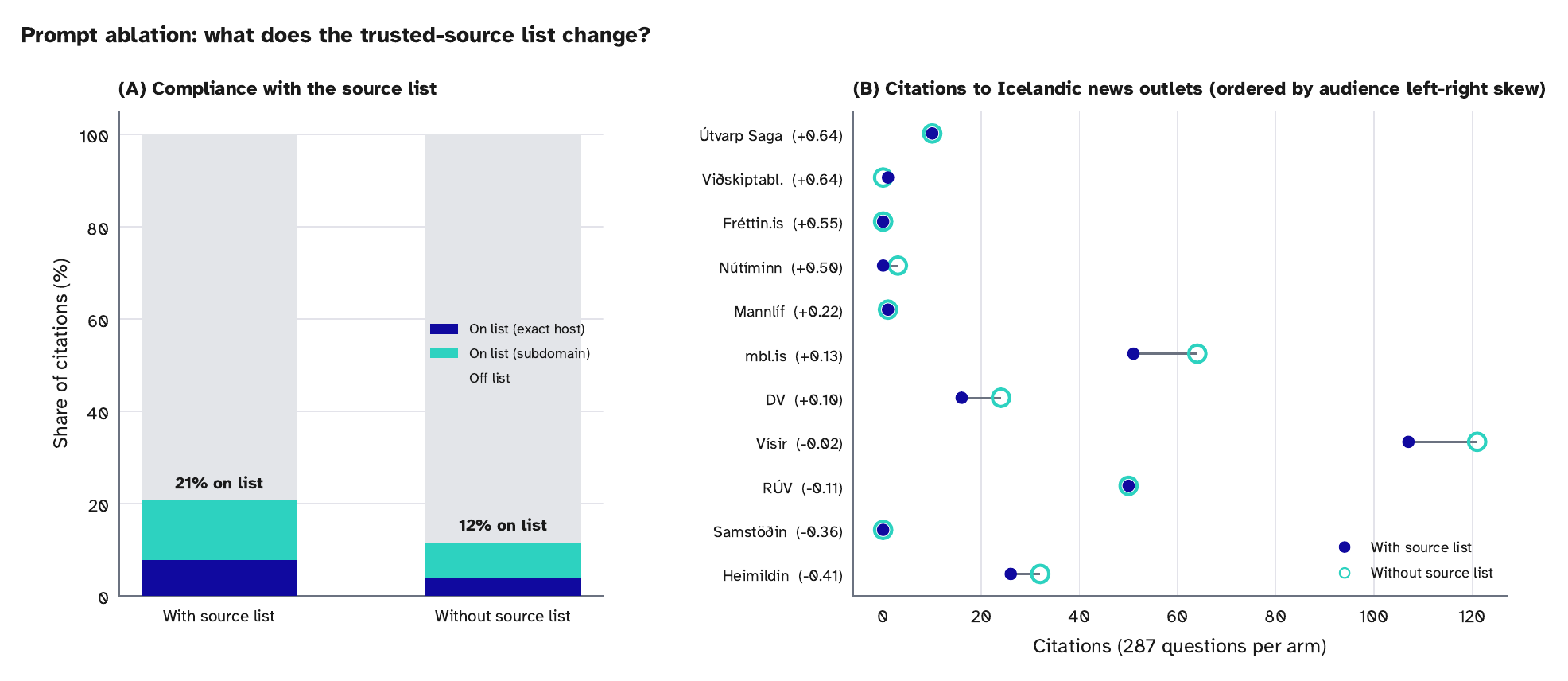}
  \caption{Prompt ablation: the 287 study questions answered with and without
  the trusted-domain list in the web-search prompt (Gemini~3 Pro, production
  API route and web-search plugin, structured citation output). (A) Share of citations on the list (exact host or
  subdomain) versus off it, per arm. (B) Citations to the surveyed Icelandic
  news outlets per arm, ordered by the left--right skew of each outlet's
  audience (in parentheses); filled = with list, ring = without. Equal counts
  show as a dot inside a ring (RÚV: 50 in each arm).}
  \label{fig:ablation}
\end{figure*}

The flag analysis showed that the web-search prompt's trusted-domain list did not keep weak sources out of the answers. The prompt ablation (Section~\ref{sec:methods-ablation}) measures directly how much that list steers the model. The answer is: only weakly. With the list in the prompt, 21\% of the 1{,}183 citations landed on a listed domain; with the list removed, 12\% of 1{,}330 did (Figure~\ref{fig:ablation}A). The instruction thus doubles compliance, but roughly four in five citations fall outside the list either way. A system prompt, on this evidence, nudges retrieval; it does not govern it.

The political profile of the cited outlets barely moves. Here, as in Section~\ref{sec:rq-profile}, ``the population'' is the national adult population as represented by the Fjölmiðlanefnd 2024 survey sample: an outlet's audience skew is measured against all respondents, so a citation-weighted skew equal to the population's means the cited outlets' combined readership is politically indistinguishable from the country as a whole. By that baseline the citation-weighted left--right skew of the cited Icelandic outlets matches the population in both arms ($-0.01$ with the list, $-0.01$ without; population $0.00$), and the nationalism--internationalism skew sits slightly to the internationalist side of the population in both ($+0.06$ relative in each). Removing the list mainly adds citations to the mainstream outlets that already dominate (Vísir from 107 to 121, mbl.is from 51 to 64; Figure~\ref{fig:ablation}B), and the alternative outlets the experts flagged most stay marginal in both arms (Útvarp Saga 10 citations in each, Fréttin.is none). RÚV is again conspicuous: the state broadcaster drew 50 citations in each arm, about 4\% of the total, identical with and without the list, modest for the country's most widely used news source and its main publicly funded newsroom.

The flat profile also contrasts sharply with the deployed system itself. In production, the citations leaned right of the population ($+0.17$), drew heavily on Útvarp Saga, and never included RÚV (Section~\ref{sec:rq-profile}). The ablation used the same model, the same questions, the same web-search plugin, and a near-identical prompt, yet its citations show none of this: the profile is neutral, Útvarp Saga is rare, and RÚV appears regularly. Only two things separate the runs. The ablation collected citations through a structured output format (with the search tool's redirect links resolved to their target pages), whereas production parses the reference lists the model writes into its free-text answers; and the runs took place a few days apart. That such small differences move the citation mix this much is an important finding in its own right: which sources an AI assistant cites is not a stable property of the model and prompt alone but of the entire configuration that produces the answer, so a measurement made on one configuration, including ours, does not automatically carry over to a differently invoked ``same'' system. The prompt, on the compliance evidence above, is among the weakest levers in that configuration. The ablation's citations were not expert-reviewed, so this comparison concerns citation patterns, not judged trustworthiness (Section~\ref{sec:limitations}).

\subsection{Reliability of the expert judgments}
\label{sec:rq-reliability}
We treat the agreement analysis as exploratory, because the overlap supporting it is thin. Of the 449 reviewed answers, only 82 (18\%) were scored by more than one reviewer, and pairwise overlap between reviewers ranges from 11 to 22 shared answers (Appendix~\ref{app:reliability}). This is a deliberate consequence of the review procedure, which prioritised un-reviewed items so that corpus coverage built up before reviewers doubled up (Section~\ref{sec:methods}), not an oversight; it does, however, mean the coefficients below should be read as indicative rather than confirmatory. On those doubly-reviewed answers the whole-answer judgments held up under chance correction. Agreement was strongest on appropriate scope (Gwet AC1 $=0.83$) and weakest on the overall ``publishable'' decision (AC1 $=0.29$), with most criteria in between. Kappa-based measures collapsed for the high-prevalence criteria. ``No hallucinations'' is the clearest case: reviewers gave the same verdict on 74\% of answers, yet Fleiss' kappa was negative. This is the kappa paradox. When almost every answer falls in one category, here nearly all were judged free of hallucinations, two reviewers will land on the same verdict most of the time simply by both picking the common label. Kappa treats that high baseline as agreement expected by chance and subtracts it, which leaves little room to score agreement above chance, so a few disagreements can push the coefficient to zero or below even when raw agreement is high. Gwet's AC1 estimates the chance baseline in a way that does not balloon under skewed prevalence, so it does not collapse, which is why we read agreement through it \citep{gwet2008}. We report the full comparison in Appendix~\ref{app:reliability}.

The source flags, which carry the headline result, have even thinner support for an agreement estimate. Only 40 of the 128 flags fall on a doubly-reviewed answer, and on just six answers did two reviewers independently flag a source, too few to estimate flag reliability. We therefore treat flags as individual expert judgments, not adjudicated rulings, and read the headline accordingly: more than a third of reviewed web answers cited a source that \emph{a reviewer} judged untrustworthy or irrelevant.

%======================================================================
\section{Discussion}
%======================================================================
\subsection{Source trustworthiness as a neglected dimension of information quality}
Our central observation is simple. The sources behind a public AI service can be assessed, they vary in trustworthiness, and that variation is rarely surfaced. Services routinely log queries and answers, but recording whether the cited material would survive expert review is not, as far as we have seen, standard practice in deployed public services. Classic information-quality work is useful here. \citet{wang1996} treat quality as something data consumers judge along many dimensions at once and group them into four families: intrinsic quality (accuracy, objectivity, believability, reputation), contextual quality (relevance, timeliness, completeness, appropriate amount), representational quality, and accessibility. The trustworthiness of a cited source belongs to the intrinsic family, close to believability and reputation, and our data show it behaving largely as a dimension of its own. Within the web path, answers with flagged sources were no worse on fluency, topicality, or scope; only a reviewer's overall publishability judgment picked up part of the problem. A quality process that scores the surface of an answer will therefore miss much of the source-trust problem, even where a careful overall editorial call catches some of it.

\subsection{Coverage and trust pull in opposite directions}
The two paths traded off along a single axis. Open web search reached current, on-point material, which is why it answered the question far more often, but much of that material came from alternative, opinion-driven outlets rather than edited mainstream journalism, and in more than a third of the answers reviewers examined, a cited web source was flagged. That happened despite an explicit instruction to prefer reputable, primary, and authoritative sources and to avoid blogs, opinion columns, and partisan sites (Appendix~\ref{app:prompts}): the flag rate shows that prompt-level steering, on its own, did not keep weak sources out of the answers, and the prompt ablation quantifies how weak that lever is, with the trusted-domain list raising the share of citations to listed domains only from 12\% to 21\% (Section~\ref{sec:rq-ablation}). Independent survey data reinforce the point: the outlets the system leaned on most heavily, and that experts flagged most, are also those the public rates as least politically independent and whose audiences sit furthest to the right and toward the nationalist pole (Section~\ref{sec:rq-profile}). The mirror image of that reliance is as striking: RÚV, the public broadcaster, the country's most widely used news source, and the outlet the surveyed public rates most politically independent, was never cited in any of the 287 web-search answers. A citation mix can thus be skewed in two directions at once, toward weak sources and away from the strongest ones, and only the first of these shows up in a flag count. The curated corpus had the mirror-image profile: its sources were trusted by construction, but it could not cover every question, and when it fell short the model declined to provide an answer rather than reaching for one. That decline is worth dwelling on. Faced with a question the corpus did not address, the model did not improvise; it said the material was not there. The rubric scores this as a failure to answer, yet it is close to what one wants from a public service: an honest ``I cannot answer that from my sources'' rather than a confident guess. Read this way, RAG's low quality score measures the coverage of the curated base more than the competence of the model, and the response is to widen and maintain that base rather than to loosen the model's grounding. Its one recurring source fault, when it did answer, was staleness. The two paths are not as separate as the design implies: in several web answers reviewers noted that a higher-quality Icelandic source, an Evrópuvefur article, existed but was not the one the model reached for. (Evrópuvefur's account of the Maastricht criteria, for instance, is authoritative enough to be cited directly in a recent University of Iceland economics report \citep{ioes2026maastricht}.) We had deliberately kept the service's own corpus off the source guidance given to the web-search mode (Section~\ref{sec:methods}), precisely so that the evaluation would measure the trustworthiness of the sources the system cited outside that domain. In production, however, adding it, so that open-web answers can fall back on vetted in-house material, is a concrete, low-cost way to push the trade-off toward trust without sacrificing coverage. Neither path is safe by default. A curated base buys trust at the price of coverage and needs constant maintenance; an open-web path buys coverage at the price of trust and needs some way to tell a reliable page from an unreliable one.

\subsection{Responses and their trade-offs}
What follows for practice is less obvious than it may seem, and we resist a single prescription. The most direct response, restricting a public service to an approved list of sources, trades one risk for another. It improves reliability but concentrates editorial power in whoever maintains the list, an uncomfortable position for a public body and one in tension with open access to information \citep{ananny2018}. A lighter-touch alternative is transparency about provenance: making the cited sources, and the basis for citing them, visible to the reader rather than buried. Content-provenance standards such as Content Credentials (C2PA) offer one technical route to carry and show where material came from \citep{c2pa2025}. Independent or expert labelling of source trustworthiness, kept separate from the institution that answers, is one way to surface that signal without handing any single actor a veto. Two prior questions complicate all of this. The first is definitional: what makes a source trustworthy is itself contested, and any labelling scheme must answer it before it can be applied. The second is jurisdictional: it is not obvious that an AI provider should be adjudicating source trust at all, since aggressive scrutiny shades into editorial control. A lighter option keeps the choice with the reader, letting a user constrain, before asking, which domains or kinds of source a query may draw on, though few mainstream systems expose such a control today. Human review, as practised here, catches both failure modes but does not scale to live answering. Keeping curated corpora current addresses staleness but not web trust. These are options with trade-offs, not a solution. Our claim is narrower: the trustworthiness of cited sources should be measured and disclosed, and the choice among responses is a matter for public deliberation.

\subsection{Implications}
For theory, the results argue for treating source trustworthiness as a first-class dimension of information quality in AI-mediated public information, alongside accuracy and currency. For practice, they point to concrete levers: log and audit the sources a service cites, report source-trust metrics as part of routine evaluation, and disclose provenance to users. For procurement and oversight, a public body adopting an LLM service should ask not only how accurate it is but where its answers come from and how that is checked.

The stakes are highest around elections, and the supply side of the problem is getting cheaper. LLMs can now produce election misinformation that human readers cannot reliably distinguish from authentic content, at high volume and negligible cost \citep{williams2025}, and analysts have warned since the models' early days that they would lower the price of large-scale influence operations \citep{goldstein2023}. Astroturfing, coordinated campaigns posing as ordinary citizens, long predates these models \citep{keller2020}, but generative AI removes its main bottleneck: the writing. The concern for a service like the one studied here is not only that voters read such material directly, but that a web-grounded assistant retrieves and cites it. That pathway is no longer hypothetical. A 2025 audit found that a Kremlin-linked network of sites, publishing millions of low-readership articles a year seemingly aimed at machines rather than people, had its claims repeated by leading chatbots in a third of tested responses \citep{newsguard2025pravda}. Follow-up research attributes such citations less to wholesale poisoning than to data voids, niche questions on which authoritative coverage is scarce \citep{alyukov2025}, which is precisely the condition of a small language during a contested referendum: on many of our questions, the trustworthy Icelandic material for a retrieval system to find is thin or absent. A referendum can plausibly be targeted by flooding the open web with material that an AI assistant will then retrieve, cite, and lend an institution's credibility to. Measuring the trustworthiness of cited sources, as done here, is therefore not only a quality-assurance exercise; it is a basic defence for the integrity of the information environment around a vote.

\subsection{Future work}
\label{sec:futurework}
Four directions follow directly from this study. First, the source-flagging instrument we used is labour-intensive and does not scale to live answering; a natural next step is an \emph{automated source-trust vetter} that surfaces weak sources (to an editor or to the reader) at answer time rather than in a retrospective audit. This need not involve any model training: an LLM, given clear and explicit criteria for what makes a source trustworthy, could vet each cited page on the fly, and our flag corpus, with its per-source labels and free-text reasons, offers both a way to specify those criteria and a benchmark to evaluate the vetter against. The main caveat is cost. Running an independent vetter alongside the answering model adds a second pass to every query, but in a public setting, where the credibility of the answers is the whole point, that added cost may well be worth paying. Second, the review instrument should add an answer-level \emph{balance} criterion. Per-source flags cannot see one-sidedness: every cited source can pass individually while the answer as a whole presents one side of a contested question. Our flag comments suggest the risk is real, since opinion and editorial pieces were among the reviewers' most common objections (Section~\ref{sec:taxonomy}), and an answer resting on a single such source can editorialise without tripping any per-source check; \citet{wang1996} would place balance under objectivity, in the same intrinsic-quality family as believability. For a service whose mandate is even-handedness, balance deserves a criterion of its own. Third, our remedies (provenance disclosure, source labelling, user-set domain constraints) are so far argued rather than tested; a \emph{user-facing study} could measure how such provenance signals and controls actually affect users' trust, verification behaviour, and reliance, which is what ultimately determines whether transparency helps. Fourth, the design should be \emph{replicated} in other low-resource languages and other high-stakes civic moments (elections, public-health emergencies), both to test how far the coverage--trust trade-off generalises and to support the larger, balanced, planned-overlap reviewer design that a confirmatory agreement estimate would require.

%======================================================================
\section{Limitations}
\label{sec:limitations}
%======================================================================
Several limitations bound these results. The evaluation was conducted before public release, on the system as configured during testing, and the questions were generated to span the debate rather than drawn from real users, so they may over- or under-represent what citizens would actually ask. Evaluations were not randomly assigned to modes: reviewers worked through a shared queue rather than a balanced design, so the two modes were reviewed in unequal numbers (262 RAG versus 187 web answers) and the mode comparison is observational rather than controlled. Because the flagging instrument applied different reasons to each retrieval path, cross-mode flag comparisons are descriptive, and the web-only flag rate (35\%) is best read on its own rather than as a like-for-like contrast with RAG. The study covers a single service, language, and topic during one referendum period, which limits generalisation. The 128 flags are a standing backlog of open flags rather than a closed, adjudicated review, so they record what reviewers raised, not a final ruling; most were single-reviewer judgments, with too little overlap to estimate their reliability. The reviewer pool is small (five experts), and inter-rater overlap is modest. Our reviewers were subject-matter experts in EU affairs and may hold priors on the question, which could be perceived as pro-EU; we mitigated this by asking them to judge cited sources against editorial criteria (trustworthiness, relevance, and currency) rather than the political position of an answer (the rubric contains no item rewarding agreement with a conclusion), and by having them work independently and unable to see one another's evaluations, but we cannot rule out that judgments of which sources count as trustworthy carry some such priors. A related risk attaches to \texttt{esbvaktin.is}: because it is itself an LLM-assisted aggregator and supplied both our question seeds and the source guidance for web-search mode, it could in principle propagate its own selection biases into our pipeline (an LLM feedback loop). We limited this by using it only as a source map, never as content for answers, and by having reviewers judge the actual cited pages; a residual risk remains that what we call ``trusted'' inherits its editorial judgments. Because judging a source means seeing it, reviewers could tell a local article from an external web page, so the retrieval mode was not blinded and may have coloured some judgments. Finally, scanning seven criteria invites multiple comparisons, so single-criterion $p$-values should be read as exploratory. The survey-based political profiling of cited outlets (Section~\ref{sec:rq-profile}) characterises each outlet's \emph{audience}, not the slant of the specific article cited, covers only the Icelandic-news subset of cited sources, and applies a 2024 survey to a system evaluated in 2026. The prompt ablation (Section~\ref{sec:rq-ablation}) used the same model and web-search plugin as the deployed system but elicited citations through a structured-output schema rather than the free-text reference lists of production answers, and ran a few days later; its two arms are internally comparable, but it measures list compliance rather than trustworthiness, and its contrast with the deployed system's citation mix confounds elicitation format with timing, so we read that contrast as evidence of configuration sensitivity rather than as a like-for-like comparison. AI capabilities also change quickly: these findings describe the specific models and the period we studied and should not be read as fixed properties of the technology. We expect the measurement approach, more than any single measurement, to carry over.

%======================================================================
\section{Conclusion}
%======================================================================
Public institutions are adopting AI to answer citizens directly, and the sources behind those answers are largely invisible, to users and often to the institutions themselves. Evaluating an independent, government-funded service before its public launch, one answering EU-related questions ahead of a national referendum, expert reviewers flagged a cited source in more than a third of the web-search answers they examined, almost always as untrustworthy or irrelevant. The trust problem was concentrated in the open-web path, and the trustworthiness of a source could not be read off the quality of the answer: fluent, on-topic answers rested on sources experts would not endorse, while the country's most widely used and most trusted news source went entirely uncited. We do not claim a single remedy. We do claim that source trustworthiness is measurable, that it matters most where public stakes are highest, and that public AI services should measure and disclose it rather than assume it.

%======================================================================
% Declarations
%======================================================================
\section*{CRediT authorship contribution statement}
\textbf{Hafsteinn Einarsson:} Conceptualization, Methodology, Software, Formal
analysis, Data curation, Writing -- original draft, Writing -- review \& editing,
Supervision. \textbf{Hafsteinn Birgir Einarsson:} Validation, Writing -- review \& editing. \textbf{Jón Gunnar Þorsteinsson:}
Supervision of reviewers, Validation, Writing -- review \& editing. \textbf{Jón Gunnar
Ólafsson:} Validation, Data Curation, Writing -- review \& editing.

\section*{Declaration of competing interest}
The authors declare that they have no known competing financial interests or
personal relationships that could have appeared to influence the work reported in
this paper.

\section*{Funding}
This work was funded by the Ministry for Foreign Affairs of Iceland. The funder
had no role in the study's design, conduct, analysis, interpretation, or
reporting, or in the decision to submit the work for publication.

\section*{Data availability}
The anonymised evaluation data (reviewer ratings and source flags) and the
analysis code used to produce the figures and statistics are available from the
corresponding author on reasonable request.

\section*{Acknowledgements}
We thank the expert reviewers who evaluated the service's answers; their
contribution was supported through funding from the Ministry for Foreign Affairs
of Iceland. We would also like to thank Professor Maximilian Conrad, School of
Social Sciences at the University of Iceland, for assisting in locating the
expert reviewers, and Brynjólfur Gauti Guðrúnar Jónsson, doctoral student in
statistics at the University of Iceland, who created \texttt{esbvaktin.is}, the
fact-checking project that supplied our source material and the trusted-domain
classification used in web-search mode. We thank Vilborg Ása Guðjónsdóttir for
helpful discussions during the writing of this manuscript, and the Icelandic
Media Commission (Fjölmiðlanefnd) for access to their 2024 survey findings.

\section*{Declaration of generative AI use}
The authors used generative AI tools to assist with editing, under author supervision. The authors reviewed and edited all content and take full responsibility for it.

%======================================================================
\appendix
%======================================================================
\section{Inter-rater reliability}
\label{app:reliability}
Of the 449 reviewed answers, 82 (18\%) were scored by more than one reviewer.
Pairwise overlap between the five reviewers is modest, ranging from 11 to 22
commonly-scored answers per pair, so Table~\ref{tab:reliability} should be read
as exploratory. The kappa paradox is visible in the high-prevalence criteria:
``no hallucinations'' has 74\% raw agreement but a negative Fleiss' $\kappa$,
because almost all judgments fall in one category. Gwet's AC1 is stable under
that prevalence and is our headline measure.

\begin{table}[t]
  \centering
  \footnotesize
  \caption{Chance-corrected agreement per criterion on the 82 doubly-reviewed
  answers. AC1 = Gwet's AC1; $\kappa$ = Fleiss' kappa; $\alpha$ = Krippendorff's
  alpha.}
  \label{tab:reliability}
  \begin{tabular}{lrrrr}
    \toprule
    Criterion & \% agree & AC1 & Fleiss' $\kappa$ & Kripp.\ $\alpha$ \\
    \midrule
    Answers question        & 65.9 & 0.39 & 0.23 & 0.25 \\
    Factually accurate      & 75.6 & 0.61 & 0.31 & 0.36 \\
    Sources relevant        & 75.6 & 0.53 & 0.49 & 0.52 \\
    No hallucinations       & 73.5 & 0.66 & $-0.21$ & $-0.14$ \\
    Appropriate scope       & 87.8 & 0.83 & 0.50 & 0.55 \\
    Language quality        & 77.2 & 0.69 & 0.14 & 0.13 \\
    Publishable, minor edits& 63.0 & 0.29 & 0.22 & 0.22 \\
    \bottomrule
  \end{tabular}
\end{table}

\section{Answer-generation prompts}
\label{app:prompts}
Both modes share the same task but differ in how they are grounded. The RAG
system prompt is reproduced first, translated from the Icelandic original as
deployed, then the web-search prompt (English in the original), reproduced with
its complete trusted-domain list. The prompt asks the model to prefer this list,
which \texttt{esbvaktin.is} classifies as ``high confidence'', but does not
enforce it as a hard constraint; the examples named in the main text
(\texttt{icelandmonitor.mbl.is} and others) are illustrative entries from it, not
a preference ordering. We deliberately kept the service's own corpus
(\texttt{evropuvefur.is}) off the list, so that the evaluation would measure the
trustworthiness of the sources the system cited outside that domain; its sister
site \texttt{visindavefur.is} does appear on the list.

\subsection*{RAG mode (system prompt, translated from Icelandic)}
{\footnotesize
\begin{verbatim}
You are an expert assistant for Evrópuvefurinn (evropuvefur.is), run by the
University of Iceland. You are an expert on European Union affairs, the European
Economic Area (EEA), and Iceland's relations with Europe.

## Scope
Answer only questions about Europe, the EU, the EEA, European integration, and
Iceland's relations with Europe. Politely decline questions that are clearly out
of scope with a single sentence pointing the user in the right direction.

## Basis for answers
Base your answers on the accompanying articles from Evrópuvefur's knowledge base.
Do not make up facts. If the context does not contain enough information to
answer, say so.

## Citations
Each article you are given is numbered. Use only markdown links of the form
[[N]](source_url). Do not invent numbers that are not among the articles given.

## Language
Answer in the same language as the user's question.

## Style and length
Write clearly, accessibly, and in an academic tone. Typical length: 150-400 words.
\end{verbatim}
}

\subsection*{Web-search mode (system prompt, full source list)}
{\footnotesize
\begin{verbatim}
You are an expert assistant for Evrópuvefurinn (evropuvefur.is), run by the
University of Iceland. You specialize in EU affairs, the EEA, and Iceland's
relations with Europe.

## Web Search Mode
You have access to web search. Use it to find current, accurate information to
answer the user's question.

## Source selection
Prioritize sources from the list below. These are the domains that esbvaktin.is
-- Iceland's EU referendum fact-checking project (University of Iceland) --
classifies as "high confidence" primary or authoritative secondary sources.
Prefer them over blogs, opinion columns, social media, aggregators, or
unfamiliar outlets. Other reputable sources may be used when they add material
the trusted list does not cover, but say so in context and avoid partisan
websites (political parties, advocacy groups) as evidence for factual claims --
cite them only for documenting a party or group's own position.

### EU institutions and legal texts
eur-lex.europa.eu, ec.europa.eu, commission.europa.eu, europarl.europa.eu,
consilium.europa.eu, europa.eu, eeas.europa.eu, enlargement.ec.europa.eu,
neighbourhood-enlargement.ec.europa.eu, agriculture.ec.europa.eu,
climate.ec.europa.eu, oceans-and-fisheries.ec.europa.eu,
taxation-customs.ec.europa.eu, cohesiondata.ec.europa.eu, stecf.ec.europa.eu,
eea.europa.eu, ecb.europa.eu, sdw.ecb.europa.eu,
ireland.representation.ec.europa.eu.

### EEA / EFTA
efta.int, eftasurv.int, eftacourt.int.

### Icelandic government, parliament, and institutions
althingi.is, stjornarradid.is, government.is, island.is, sedlabanki.is, cb.is,
orkustofnun.is, mast.is, skatturinn.is, rna.is, ust.is, fjolmidlanefnd.is,
ferdamalastofa.is, stjornlagarad.is.

### Icelandic statistics and academic reference
hagstofa.is, px.hagstofa.is, statice.is, visindavefur.is, uni.hi.is.

### International organizations
oecd.org, data-explorer.oecd.org, imf.org, data.worldbank.org, fao.org, bis.org,
nato.int, coe.int, hdr.undp.org, eeagrants.org, ccpi.org.

### Foreign governments, parliaments, and agencies
gov.uk, legislation.gov.uk, commonslibrary.parliament.uk,
researchbriefings.files.parliament.uk, obr.uk, gov.ie, regeringen.dk, fm.dk,
thedanishparliament.dk, lf.dk, riksdagen.se, ei.se, su.se, stat.fi, mmm.fi,
ssb.no, fiskeridir.no, europa.eda.admin.ch, hnb.hr.

### Icelandic social partners, industry, and polling
sa.is, asi.is, si.is, bondi.is, responsiblefisheries.is, islandsbanki.is,
gallup.is, northstack.is, eurometal.net.

### Scholarly, legal, and reference
doi.org, jstor.org, link.springer.com, avalon.law.yale.edu,
constituteproject.org, en.wikipedia.org (for orientation only -- always prefer
primary sources for citations).

### Reputable news outlets
reuters.com, politico.eu, rte.ie, theskipper.ie, thelocal.com,
nordiclabourjournal.org, grapevine.is, opendemocracy.net, icelandmonitor.mbl.is,
europeanmovement.ie, arcticcircle.org.

## Citations
Use numbered-bracket citations with inline links: [[N]](URL). Assign numbers in
order of first appearance and reuse them. After the answer, add a References
section listing each cited source. Do not invent sources or URLs; only cite
sources you actually retrieved via web search.
\end{verbatim}
}

\section{Flag-comment coding audit}
\label{app:flagaudit}
Table~\ref{tab:flagaudit} lists all 77 reviewer comments on flagged web sources,
an English translation, and the reason codes assigned by the LLM and reviewed by
the authors. It lets the reader check the coding against the original evidence.

\begin{footnotesize}
\begin{longtable}{p{0.36\linewidth}p{0.40\linewidth}p{0.18\linewidth}}
\caption{All 77 reviewer comments on flagged web sources, with an English translation and the reason codes assigned by the LLM and reviewed by the authors.\label{tab:flagaudit}}\\
\toprule
Comment (Icelandic) & Translation (English) & Reason codes \\
\midrule
\endfirsthead
\multicolumn{3}{c}{\tablename\ \thetable\ -- continued}\\
\toprule
Comment (Icelandic) & Translation (English) & Reason codes \\
\midrule
\endhead
\bottomrule
\endfoot
Heimild á öðru tungumáli en íslensku og ensku. & Source in a language other than Icelandic and English. & foreign language \\
\addlinespace[2pt]
Vafasöm heimild sem er ekki frá nógu vönduðum miðill, rithöfundur hefur viðurkennir að hann notar gervigreind við skrif. & Questionable source that is not from a sufficiently reputable outlet; the author has admitted to using artificial intelligence in his writing. & poor quality \\
\addlinespace[2pt]
Þetta er bloggsíða sem er ekki fræðilega traust. & This is a blog website that is not academically sound. & blog \\
\addlinespace[2pt]
Þessi grein er einnig ekki fræðilega hlutlaus. & This article is also not theoretically neutral. & partisan source \\
\addlinespace[2pt]
Þetta er hlutdræg skoðanagrein sem er ekki fræðilega traust. & This is a biased opinion piece that is not theoretically sound. & opinion piece, partisan source \\
\addlinespace[2pt]
Ekki traustvekjandi miðill að mínu mati & Not a trustworthy medium in my opinion & other \\
\addlinespace[2pt]
Þó svo að Björn sé virtur stjórnmálamaður að þá er hann ekki að notast við neinar heimildir í skrifum sínum og því væri hægt að nota betri heimild hér. & Even though Björn is a respected politician, he does not use any sources in his writings, and therefore a better source could be used here. & no references \\
\addlinespace[2pt]
Vil frumheimild! & Want the primary source! & cite primary \\
\addlinespace[2pt]
Hér væri hægt að nota betri heimild, greinin frekar stutt og vísar ekki í heimildir. Svarið við þessari spurningu snýst ekki um skoðanir heldur staðreyndir og því fagmannlegra að hafa peer-reviewed greinar sem heimildir & A better source could be used here; the article is rather short and does not cite sources. The answer to this question is not about opinions but facts, and therefore it is more professional to use peer-reviewed articles as sources. & no references, cite primary \\
\addlinespace[2pt]
Vil frumheimild & Want original source & cite primary \\
\addlinespace[2pt]
Vil frumheimild & Want the primary source & cite primary \\
\addlinespace[2pt]
Þetta er skoðanagrein sem er líklega ekki fræðilega traust. & This is an opinion piece that is probably not academically sound. & opinion piece \\
\addlinespace[2pt]
Þetta er skoðanagrein sem er ekki fræðilega traust. & This is an opinion piece that is not academically robust. & opinion piece \\
\addlinespace[2pt]
Þessi grein er ekki hlutlaus (birt af Evrópuhreyfingunni). & This article is not neutral (published by the European Movement). & partisan source \\
\addlinespace[2pt]
ályktun svarsins byggir á þessu svari sem er hlutdrægt. & The conclusion of the response is based on this answer, which is biased. & partisan source \\
\addlinespace[2pt]
Þetta er líka skoðanagrein sem tekin er af hlutdrægum miðli. Betra væri að vísa beint í þær sérfræðingaskýrslur sem höfundur greinarinnar nefnir. & This is also an opinion piece taken from a biased source. It would be better to refer directly to the expert reports mentioned by the author of the article. & opinion piece, partisan source, cite primary \\
\addlinespace[2pt]
Þetta er skoðanagrein sem er ekki hlutlaus og ekki fræðilega traust. & This is an opinion piece that is not objective and lacks academic rigor. & opinion piece, partisan source \\
\addlinespace[2pt]
Bæði ekki traustur miðill og svo hefur þjóðaratkvæðagreiðsla verið staðfest 29. ágúst. & Both not a reliable source, and also a referendum has been confirmed for August 29th. & outdated \\
\addlinespace[2pt]
Óaðgengileg heimild, það þarf áskrift til að lesa greinina. & Inaccessible source, a subscription is required to read the article. & paywalled \\
\addlinespace[2pt]
Wikipedia heimild á öðru tungumáli. Mjög erfitt að sannreyna þar sem síðurnar eru ekki alltaf með sömu upplýsingar á mismunandi tunugmálum. & Wikipedia source in another language. Very difficult to verify since the pages do not always have the same information in different languages. & Wikipedia, foreign language \\
\addlinespace[2pt]
Þessi heimild er erfitt að rekja, þetta er bara pdf sem hefur engar upplýsingar um útgefanda, blað né netsíðu og hefur enga heimildaskrá. & This source is difficult to trace; it is just a PDF that has no information about the publisher, journal, or website, and has no bibliography. & unknown outlet, no references \\
\addlinespace[2pt]
Saga er ekki hlutlausasta heimildinn (gagnvart ESB) til þess að vísa í svo það er kannski ekki nógu áreiðanleg heimild. & Saga is not the most neutral source (regarding the EU) to refer to, so it is perhaps not a reliable enough source. & partisan source \\
\addlinespace[2pt]
Wikipedia er kannski ekki nógu áreiðanleg heimild þar sem hægt er að breyta gögnunum þar inni án þess að taka það fram. & Wikipedia is perhaps not a reliable enough source, as the data in there can be edited without it being explicitly stated. & Wikipedia \\
\addlinespace[2pt]
Þetta er skoðanagrein sem er ekki hlutlaus og ekki fræðilega traust. & This is an opinion piece that is not neutral and not academically sound. & opinion piece, partisan source \\
\addlinespace[2pt]
Virkar ekki!!! & Does not work!!! & broken link \\
\addlinespace[2pt]
óþarfi og óáreyðanleg í þessu samhengi. & unnecessary and unreliable in this context. & irrelevant, other \\
\addlinespace[2pt]
Wikipedia er ekki áreiðanleg heimild þar sem hver sem getur skrifað og breytt gögnum, auk þess er heimildnotkun á íslensku útgáfunni ekki alltaf vel ritskoðað. & Wikipedia is not a reliable source since anyone can write and edit data; furthermore, the use of sources in the Icelandic version is not always well-reviewed. & Wikipedia \\
\addlinespace[2pt]
Almennt er ekki talið viðeigandi að nota Wikipedia sem heimild þar sem einstaklingar geta ávallt breytt upplýsingunum. & Generally, it is not considered appropriate to use Wikipedia as a source since individuals can always change the information. & Wikipedia \\
\addlinespace[2pt]
Þetta virðist aðeins utan fyrir efni spurningarinnar. & This seems slightly off-topic for the question. & irrelevant \\
\addlinespace[2pt]
Þetta virðist vera persónuleg blogsíða. Hún vísar í mikla tölfræði og góð gögn en vísar ekki í neinar heimildir, en er skrifuð af sérfræðingi með mikla reynslu, svo það er matsatriði hvort þetta sé nógu traustverðug heimild. & This seems to be a personal blog. It refers to a lot of statistics and good data but does not cite any sources; however, it is written by an expert with extensive experience, so it is a matter of judgment whether this is a sufficiently reliable source. & blog, no references \\
\addlinespace[2pt]
Flaggaði þessa heimild annarsstaðar þar sem talað var um staðreyndir en hún er vel viðeigandi hér þar sem er verið að tala um hver umræðan í samfélaginu sé um þessa fullyrðingu & I flagged this source elsewhere where facts were being discussed, but it is very appropriate here since the topic is what the public debate is regarding this claim. & other \\
\addlinespace[2pt]
Þetta er skoðanagrein sem er ekki fræðilega traust. & This is an opinion piece that is not theoretically sound. & opinion piece \\
\addlinespace[2pt]
Þetta kemur af bloggsíðu sem er ekki hlutlaus og ekki fræðilega traust. & This comes from a blog that is not neutral and not academically reliable. & blog, partisan source \\
\addlinespace[2pt]
Þetta er skoðanagrein sem er ekki hlutlaus og ekki fræðilega traust. & This is an opinion piece that is not objective and lacks academic rigor. & opinion piece, partisan source \\
\addlinespace[2pt]
Þetta er skoðanagrein sem hentar ekki fyrir fræðilega umfjöllun. & This is an opinion piece that is not suitable for academic discussion. & opinion piece \\
\addlinespace[2pt]
Þetta virðist vera aðsend grein á netmiðli sem notar ekki heimildaskrá, svo mér finnst hún ekki nógu traustverðug. & This seems to be a contributed article on an online outlet that does not use a bibliography, so I do not find it trustworthy enough. & opinion piece, no references \\
\addlinespace[2pt]
Ekki áreiðanleg heimild. Þessi heimild er frá sjálfstæðu netblaði sem opinberlega nýtur gervigreind við skrif fréttagreina og virðist aðallega hafa einn höfund svo það virðist ekki hafa nógu góða ritstjórn. Þessi tiltekna grein var leiðrétt fyrir að hafa rangar staðhæfingar svo það er ekki viðeigandi að nota þetta sem heimild. & Not a reliable source. This source is from an independent online newspaper that openly uses artificial intelligence to write news articles and appears to have mostly a single author, so it does not seem to have sufficient editorial oversight. This particular article was corrected for containing false statements, so it is not appropriate to use it as a source. & poor quality \\
\addlinespace[2pt]
Þetta er skoðanagrein sem er ekki hlutlaust né fræðilega traust. & This is an opinion piece that is neither neutral nor academically sound. & opinion piece, partisan source \\
\addlinespace[2pt]
Þetta er skoðanagrein sem er líklega ekki hlutlaus né fræðilega traust. & This is an opinion piece that is likely neither neutral nor academically sound. & opinion piece, partisan source \\
\addlinespace[2pt]
Það er til betri heimild um Maastricht-skilyrðin á Evrópuvefnum. & There is a better source on the Maastricht criteria on Evrópuvefurinn. & cite primary \\
\addlinespace[2pt]
Ekki nota blogg, ekki áreiðanlegt. Frekar akademíska texta, greinar í virtum blöðum etc. & Do not use blogs, they are not reliable. Rather use academic texts, articles in reputable journals, etc. & blog \\
\addlinespace[2pt]
Ekki hægt að tala um skoðanir ungs fólks þegar þær koma ekki fram hér. Aðeins fjallað um skoðanir fullorðinna í Framsókn á Norðurlandi og Lilja Alfreðsdóttir er frá Reykjavík svo hún getur ekki talað fyrir Framsókn út á landi. & It is not possible to talk about the views of young people when they are not presented here. It only discusses the views of adults in the Progressive Party in the North, and Lilja Alfreðsdóttir is from Reykjavík, so she cannot speak for the Progressive Party in the countryside. & irrelevant \\
\addlinespace[2pt]
Ekki viðeigandi þar sem hér er hvorki talað um Framsókn né við Framsóknarmann & Not applicable since there is neither mention of Framsókn (the Progressive Party) here, nor is this addressed to a member of Framsókn. & irrelevant \\
\addlinespace[2pt]
Væri betra að vitna beint í þessi orð framkvæmdastjóra Deutsche Bank. & Would it be better to directly quote these words from the CEO of Deutsche Bank? & cite primary \\
\addlinespace[2pt]
Ekki nægilega vel skrifuð grein, skringileg uppsett, sem er alltaf merki um óáreiðanleika & Not a well-enough written article, strangely formatted, which is always a sign of unreliability. & poor quality \\
\addlinespace[2pt]
Sama hér & Same here & other \\
\addlinespace[2pt]
Ekki Wikipedia takk, allir og amma þeirra sem geta skrifað inn á hana. & No Wikipedia please, everyone and their grandmother can write on it. & Wikipedia \\
\addlinespace[2pt]
Lesandi þarf áskrift til að lesa greininga, er það vandamál mögulega? & The reader needs a subscription to read the analysis, could that potentially be a problem? & paywalled \\
\addlinespace[2pt]
Frumheimild hér, alltof skoðunarglaður hér. Já, atkvæðagreiðslan snýst um hvort umræður eiga að hefjast að nýju en ef það á að hefja aftur umræður er planið að fara í Evrópusambandið. Þú sækir ekki um starf og ferð í 5 atvinnuviðtöl ef þú ætlar þér ekki að taka starfinu ef það býðst þér. & Primary source here, way too opinionated here. Yes, the vote is about whether to resume talks, but if talks are to be resumed, the plan is to join the European Union. You don't apply for a job and go to 5 interviews if you don't intend to take the job if it's offered to you. & opinion piece, cite primary \\
\addlinespace[2pt]
Hér á gervigreindin að nota upprunalegu heimildina sem er grein Hannesar Hólmsteins þar sem í þessari frétt er greinin einungis lauslega þýdd. & Here, the AI should use the original source, which is Hannes Hólmsteinn's article, as this news piece is only a loose translation of that article. & cite primary \\
\addlinespace[2pt]
Þessi skoðanagrein er ekki hlutlaus og ekki fræðilega traust. & This opinion piece is not neutral and is not academically sound. & opinion piece, partisan source \\
\addlinespace[2pt]
Þessi heimild er ekki fræðilega traust. & This source is not academically reliable. & poor quality \\
\addlinespace[2pt]
Heimildin er ekki fræðilega traust. & The source is not academically reliable. & poor quality \\
\addlinespace[2pt]
Get ekki opnað hlekkinn. & I can't open the link. & broken link \\
\addlinespace[2pt]
Upprunaleg heimild af bloggi, ekki viðeigandi hér & Original source from a blog, not appropriate here & blog, irrelevant \\
\addlinespace[2pt]
Ekki nægilega áreiðanleg þar sem höfundur vísar ekki í heimildir. Vil sjá betri heimild hér, helst íslenska. & Not sufficiently reliable as the author does not cite sources. I want to see a better source here, preferably an Icelandic one. & no references, cite primary \\
\addlinespace[2pt]
Þarf að finna greinina sem grevigreindin vísar í, kemur ekki í þessum hlekk. Ef fundið er réttan hlekk er þessi heimild í lagi & The article that the AI refers to needs to be found; it does not appear at this link. If the correct link is found, this source is fine. & broken link \\
\addlinespace[2pt]
Væri gott að vera með nýlegri heimild um afstöðu Íhaldsflokksins. & It would be good to have a more recent source on the Conservative Party's position. & outdated \\
\addlinespace[2pt]
Væri betra að finna heimild sem segir þetta beint en ekki frétt sem er unnin upp úr Facebook færslu. & It would be better to find a source that says this directly, rather than a news article based on a Facebook post. & cite primary, social media \\
\addlinespace[2pt]
Þessi heimild fjallar ekki um þetta. & This source does not deal with this. & irrelevant \\
\addlinespace[2pt]
Þessi heimild fjallar aðallega um áhrifin sem Brexit hafði á Danmörku. Trúlega til meira viðeigandi heimild sem fjallar um meira um Bretland. & This source mainly deals with the impact Brexit had on Denmark. There is probably a more relevant source that focuses more on the UK. & irrelevant \\
\addlinespace[2pt]
Veit ekki hversu traust heimild blöggfærsla fyrrum ráðherra Sjálfstæðisflokksins er. & I don't know how reliable a source a blog post by a former minister of the Independence Party is. & partisan source, blog \\
\addlinespace[2pt]
Almennt er Wikipedia ekki talin nógu traust heimild til að vitna í. & Generally, Wikipedia is not considered a reliable enough source to cite. & Wikipedia \\
\addlinespace[2pt]
Röng slóð, það þyrfti að taka frá síðasta skástrikið svo rétt síða komi upp. & Wrong URL, the trailing slash needs to be removed so the correct page loads. & broken link \\
\addlinespace[2pt]
Heimild 1 er hvorki á ensku né íslensku sem gerir erfitt að meta áreiðanleika, en auk þess er hún að tala gegn gagnrýninni á fjórfrelsinu. Það væri í lagi ef vitnað væri í þessi mótrök, en það er óæskilegt að nota þetta sem heimild um rök sem höfundur er að gagnrýna, auk þess að heimildinn vísar mjög stuttlega í þessi rök og útskýrir þau ekki að fullu. Það væri miklu áreiðanlegra að vísa í heimild sem fer ýtarlega í málið, getur útskýrt hvað þessir dómar snérust um og útskýrir ýtarlega rökin frá þessu sjónarhorni. & Source 1 is neither in English nor Icelandic, which makes it difficult to assess its reliability; furthermore, it argues against the criticism of the four freedoms. This would be acceptable if these counterarguments were being cited, but it is undesirable to use this as a source for arguments that the author is criticizing, in addition to the fact that the source refers to these arguments very briefly and does other not fully explain them. It would be much more reliable to refer to a source that goes into the matter in detail, can explain what these judgments were about, and thoroughly explains the arguments from this perspective. & foreign language, cite primary \\
\addlinespace[2pt]
Vil heimildina annarsstaðar frá, formlegri síðu/grein t.d., áreiðanlegra þegar er verið að tala um staðreyndir en hún er flott þegar það er verið að tala um skiptar skoðanir í seinustu málsgrein & I want the source from elsewhere, a more formal page/article for example, which is more reliable when discussing facts, but it is great when discussing differing opinions in the last paragraph. & opinion piece, cite primary \\
\addlinespace[2pt]
Ég hef ekki heyrt um þessa fréttasíðu áður og er lítið sem kemur upp þegar ég gúggla nafnið. Mér sýnist samt að allar upplýsingarnar séu réttar, en svipaðar upplýsingar koma líka fram hér í grein frá Evrópuvefnum sem mér þykir áreiðanlegri: https://www.evropuvefur.is/svar.php?id=63420 & I have not heard of this news site before, and very little comes up when I google the name. However, it seems to me that all the information is correct, but similar information can also be found here in an article from Evrópuvefurinn, which I consider to be more reliable: https://www.evropuvefur.is/svar.php?id=63420 & unknown outlet \\
\addlinespace[2pt]
Þetta er skoðanagrein sem er ekki fræðilega traust né ritrýnd. Auk þess er höfundur hennar yfirlýstur andstæðingur ESB og er greinin því ekki hlutlaus. & This is an opinion piece that is neither academically robust nor peer-reviewed. Furthermore, its author is an avowed opponent of the EU, and the article is therefore not unbiased. & opinion piece, partisan source \\
\addlinespace[2pt]
Mögulega úreltar upplýsingar þar sem húsnæðisverð hefur verið að lækka ólíkt því sem pistlahöfundur spáði fyrir um. & Possibly outdated information, as housing prices have been decreasing, contrary to what the columnist predicted. & outdated \\
\addlinespace[2pt]
Set spurningamerki við að nota BS-ritgerð sem trausta heimild. & I question the use of a BS thesis as a reliable source. & other \\
\addlinespace[2pt]
Sem eina heimildin sem svarið vitnar í tel ég ekki að þessi frétt sem byggir á greiningum hagsmunaðila vera nægilega traust. & As the only source cited in the answer, I do not consider this news article, which is based on analyses by stakeholders, to be sufficiently reliable. & partisan source \\
\addlinespace[2pt]
Hlekkurinn virkar ekki. & The link does not work. & broken link \\
\addlinespace[2pt]
Hlekkurinn virkar ekki. & The link does not work. & broken link \\
\addlinespace[2pt]
Þetta er bloggsíða sem er ekki ritrýnd né fræðilega traust. & This is a blog that is neither peer-reviewed nor academically reliable. & blog \\
\addlinespace[2pt]
Ég hef ekki heyrt um þessa síðu áður og finnst ólíklegt að hún sé ritrýnd og/eða áreiðanleg. & I have not heard of this website before and find it unlikely that it is peer-reviewed and/or reliable. & unknown outlet \\
\addlinespace[2pt]
Útvarp Saga er þekkt sem fréttamiðill sem er opinberlega gegn aðild að ESB, svo það er mögulega ekki alveg nógu áreiðanleg heimild og betra að vísa í heimildirnar sem fréttin er að vísa í. & Útvarp Saga is known as a news outlet that is officially opposed to EU membership, so it may not be a fully reliable source, and it is better to cite the sources that the article itself refers to. & partisan source, cite primary \\
\addlinespace[2pt]
Efnið snýst um vísindalegt málefni, og þá finnst mér það ekki nógu traustverðugt að vísa í grein um efnið á fréttasíðu sem hefur engar heimildir. & The subject matter is scientific, and therefore I do not find it credible enough to refer to an article on the topic on a news website that has no sources. & no references, cite primary \\
\addlinespace[2pt]
\end{longtable}
\end{footnotesize}

\section{Audience political profile of cited outlets}
\label{app:profile}
Table~\ref{tab:profile} reports, for each surveyed Icelandic news outlet, the
political make-up of its audience in the Fjölmiðlanefnd 2024 survey
\citep{fjolmidlanefnd2024} alongside how often the web-search mode cited it and
how many of its cited sources experts flagged (Section~\ref{sec:rq-profile}).
Audience skew is the share of an outlet's readers at the high end of a scale
minus the share at the low end; the national population sits at $+0.00$ on
left--right and $+0.22$ on nationalism--internationalism, so positive left--right
values mark a right-leaning audience and lower nationalism--internationalism
values a more nationalist one. Independence is the share of respondents with an
opinion who agree the outlet is ``independent of political interests''. Outlets
are ordered left to right by audience left--right skew.

\begin{table}[t]
  \centering
  \footnotesize
  \caption{Audience political profile (Fjölmiðlanefnd 2024), citation volume, and
  expert flags for the surveyed Icelandic news outlets. L--R = left--right
  audience skew; N--I = nationalist (lower) to internationalist (higher) audience
  skew; Indep. = \% rating the outlet politically independent; Cited = web-search
  answers citing it; Flags = expert source flags. RÚV and Mannlíf were not cited.}
  \label{tab:profile}
  \begin{tabular}{lrrrrr}
    \toprule
    Outlet & L--R & N--I & Indep. & Cited & Flags \\
    \midrule
    Heimildin      & $-0.41$ & $+0.60$ & 49\% & 29 & 0 \\
    Samstöðin      & $-0.36$ & $+0.52$ & 21\% & 4  & 0 \\
    RÚV            & $-0.11$ & $+0.37$ & 62\% & 0  & 0 \\
    Vísir          & $-0.02$ & $+0.27$ & 61\% & 33 & 8 \\
    DV             & $+0.10$ & $+0.25$ & 38\% & 12 & 1 \\
    mbl.is         & $+0.13$ & $+0.16$ & 31\% & 12 & 1 \\
    Mannlíf        & $+0.22$ & $+0.24$ & 28\% & 0  & 0 \\
    Nútíminn       & $+0.50$ & $+0.14$ & 30\% & 11 & 7 \\
    Fréttin.is     & $+0.55$ & $-0.08$ & 22\% & 11 & 3 \\
    Útvarp Saga    & $+0.64$ & $-0.20$ & 15\% & 30 & 12 \\
    Viðskiptabl.   & $+0.64$ & $+0.38$ & 30\% & 11 & 0 \\
    \bottomrule
  \end{tabular}
\end{table}

%======================================================================
\bibliographystyle{cas-model2-names}
\bibliography{cas-refs}
\end{document}